\documentclass[12pt,preprint]{aastex}
\usepackage{longtable}
\shorttitle{Testing Model Atmospheres for VLMS and BDs}
\shortauthors{Tottle \& Mohanty}
\begin{document}
\def\logg{$\log g$}
\def\del{{\bf{\nabla}} }
\def\delsq{{\nabla}^2 }
\def\deg{$^{\circ}$ }
\def\mdotd{\dot M_D }
\def\mdot{\dot M }
\def\dy{\frac{dy}{dx}} 
\def\dysq{\frac{d^2y}{dx^2}} 
\def\du{\frac{du}{dx}} 
\def\dusq{\frac{d^2u}{dx^2}} 
\def\kms{kms$^{-1}$} 
\def\ang{$\mathrm{\AA}$} 
\def\um{$\mu$m} 
\def\mic{$\mu$m} 
\def\Ha{H$\alpha$} 
\def\FHa{$F_{\mathrm{H}\alpha}$} 
\def\EWHa{$EW_{\mathrm{H}\alpha}$} 
\def\FcHa{$\mathcal{F}_{c\mathrm{H}\alpha}$} 
\def\Teff{$T_{\mathrm{eff}}$}
\def\Tc{$T_{\mathrm{c}}$} 
\def\alp{$\alpha$} 
\def\veff{$\emph{\textbf{v}}_{\mathrm{eff}}$} 
\def\LHa{$L_{\mathrm{H}\alpha}$} 
\def\Lbol{$L_{\mathrm{bol}}$} 
\def\LL{$L_{\mathrm{L}}$} 
\def\AL{$L_{\mathrm{A}}$} 
\def\LT{$T_{\mathrm{L}}$} 
\def\AT{$T_{\mathrm{A}}$} 
\def\AR{$R_{\mathrm{A}}$}
\def\LR{$R_{\mathrm{L}}$}
\def\LAJ{$A_{J,\mathrm{L}}$} 
\def\AAJ{$A_{J,\mathrm{NIR}}$} 
\def\eAD{$\eta_{\mathrm{AD}}$} 
\def\ed{$\eta_d$} 
\def\eOhm{$\eta_{\mathrm{Ohm}}$} 
\def\Lsun{$L_\odot$} 
\def\Msun{$M_\odot$\,} 
\def\Mj{$M_J$} 
\def\Rsun{$R_\odot$} 
\def\Rx{$R_x$} 
\def\Ms{$M_\star$} 
\def\Mhbmm{$m_{\mathrm{HBMM}}$} 
\def\Mj{$M_j$} 
\def\AV{$A_V$}
\def\AJ{$A_J$} 
\def\AH{$A_H$} 
\def\AK{$A_K$} 
\def\Ath{$A_{3.6}$} 
\def\Afo{$A_{4.5}$} 
\def\Afi{$A_{5.8}$} 
\def\Aei{$A_{8.0}$} 
\def\Atw{$A_{24}$} 
\def\Rv{$R_{v}$} 
\def\Rt{$R_{t}$} 
\def\Rx{$R_{x}$} 
\def\Rs{$R_{\star}$} 
\def\Rin{$R_{in}$} 
\def\Mdw{$\dot{M_{w}}$} 
\def\Mdd{$\dot{M_{D}}$} 
\def\be{\begin{equation}}
\def\ee{\end{equation}}
\def\nn{\nonumber}
\def\JH{$J$$-$$H$}
\def\HK{$H$$-$$K_s$}
\def\JK{$J$$-$$K_s$}
\def\Ks{$K_s$}
\def\Kt{$K_s$$-$$[3.6]$}
\def\tf{$[3.6]$$-$$[4.5]$}
\def\ff{$[4.5]$$-$$[5.8]$}
\def\fe{$[5.8]$$-$$[8.0]$}
\def\etf{$[8.0]$$-$$[24]$}
\def\apjs{ApJS}
\def\apj{ApJ}
\def\apjl{ApJL}
\def\aj{AJ}
\def\aap{AJ}
\def\pasp{PASP}
\def\nat{Nature}
\def\araa{ARA\&A}

\title{Testing Model Atmospheres for Young Very Low Mass Stars and Brown Dwarfs in the Infrared: Evidence for Significantly Underestimated Dust Opacities}
\author{Jonathan Tottle\altaffilmark{1}, Subhanjoy Mohanty\altaffilmark{1}}
\altaffiltext{1}{Imperial College London, 1010 Blackett Lab., Prince Consort Road, London SW7 2AZ, UK.  {\tt j.tottle11@imperial.ac.uk}}  

\begin{abstract}
We test state-of-the-art model atmospheres for young very low-mass stars and brown dwarfs in the infrared, by comparing the predicted synthetic photometry over 1.2--24\,$\mu$m to the observed photometry of M-type spectral templates in star-forming regions. We find that {\it (1)} in both early and late young M types, the model atmospheres imply effective temperatures (\Teff) several hundred Kelvin lower than predicted by the standard Pre-Main Sequence spectral type-\Teff\ conversion scale (based on theoretical {\it evolutionary} models). It is only in the mid-M types that the two temperature estimates agree. {\it (2)} The \Teff\ discrepancy in the early M types (corresponding to stellar masses $\gtrsim$\,0.4\,\Msun\,at ages of a few Myr) probably arises from remaining uncertainties in the treatment of atmospheric convection within the atmospheric models, whereas in the late M types it is likely due to an underestimation of dust opacity. {\it (3)} The empirical and model-atmosphere $J$-band bolometric corrections are both roughly flat, and similar to each other, over the M-type \Teff\ range. Thus the model atmospheres yield reasonably accurate bolometric luminosities (\Lbol), but lead to underestimations of mass and age relative to evolutionary expectations (especially in the late M types) due to lower \Teff. We demonstrate this for a large sample of young Cha I and Taurus sources. {\it (4)} The trends in the atmospheric model \JK\ colors, and their deviations from the data, are similar at Pre-Main Sequence and Main Sequence ages, suggesting that the model dust opacity errors we postulate here for young ages also apply at field ages.

\end{abstract}


\section{Introduction}
The mass of a star or brown dwarf (BD) is its most fundamental attribute, as the prime determinant of the object's interior and global properties and evolutionary path. Indeed, the very distinction between stellar and substellar objects is predicated on mass: only bodies more massive than $\sim$0.072\,\Msun\,can sustain stable hydrogen fusion, and thus qualify as stars. Moreover, an accurate determination of the frequency distribution of (sub)stellar masses at young ages, i.e., of the Initial Mass Function (IMF), is vital for discriminating between various theories of star formation. These issues are particularly relevant to spectral classes M and later -- which comprise very low mass stars (VLMS) and BDs -- for the following reasons. First, M dwarfs make up $\sim$80\% of the stars in our galaxy, so an understanding of their intrinsic properties is key to stellar astrophysics in general. This is especially germane given that planets appear to be ubiquitous around these stars \citep[e.g.,][]{Dressing13,Bonfils11}, and planetary formation, evolution and characteristics are inextricably linked to the properties of the host star. Second, the transition from stars to BDs occurs somewhere between the M and L types (depending on age). Third, VLMS and BDs evince novel interior physics (full convection, support by electron degeneracy pressure) and complex atmospheric phenomena (cool and high gravity conditions, dominance of molecular and dust opacities, cloud formation, ``weather''), the modelling of which is highly non-trivial, and requires both anchoring by, and testing against, empirically determined stellar/substellar parameters. Concurrently, the modelling of such cool and complex photospheres is critical for gaining insights into exoplanet atmospheres, which are largely unobservable with current technology. Finally, the shape of the IMF at the lowest masses provides a stringent test of star formation theories. As such, the determination of masses for VLMS and BDs, and of the relationship between their mass and other parameters such as temperature and luminosity, is vital.

The most direct estimate of mass comes from the calculation of orbital parameters in binary or higher-order multiple systems, which yields a dynamical mass. Appropriate systems, however, are very rare. In their absence, all estimates depend on theoretical models: either theoretical evolutionary tracks, or synthetic atmospheres, or some combination of the two. Unfortunately, the extreme paucity of empirical masses for young VLMS and BDs means that the evolutionary and atmospheric models for these objects are largely untested at Pre-Main Sequence (PMS) ages: precisely the regime where these models are most uncertain \citep{Baraffe02}, and also the regime where accurate masses are needed to constrain the IMF and thus formation scenarios. At the same time, in the only two PMS VLMS/BD systems with empirically determined component masses so far (the young eclipsing binaries Par1802AB and 2MASS0535AB), there are significant discrepancies between the observed masses and those predicted by the models from temperature/luminosity/spectral considerations, implying there may indeed be important flaws in the models for these ages and masses \citep{Stassun06,Stassun08,Mohanty09,Mohanty10,Mohanty12}. 

The goal of this paper is to compare the independent predictions of evolutionary and atmospheric models for PMS VLMS and BDs against each other, in order to test whether they are mutually consistent, and thereby probe which physical phenomena might be inadequately accounted for in these models. In particular, the determination of an accurate \Teff\ is one of the most model-dependent steps in estimating the mass. When using model atmospheres, one derives \Teff\ by fitting individual spectral features, or the overall spectral energy distribution (SED), with synthetic spectra \citep[e.g.,][]{Leggett01,Leggett02,Mohanty04a}. Alternatively, one uses a spectral type-to-\Teff\ conversion scale, which is devised to agree with the predictions of theoretical evolutionary models \citep[discussed further in \S2.3]{Luhman03,Luhman07}. Note that, while the evolutionary models use model atmospheres as an outer boundary condition, their predicted \Teff\ (and bolometric luminosities, \Lbol) as a function of age depend mostly on interior physics and only very weakly on the atmospheric opacity $\kappa$: \Teff$(t)$ $\propto \kappa^{\sim 1/10}$ (and \Lbol$(t)$ $\propto \kappa^{\sim 1/3}$) \citep{Burrows93}. The shape of the SED and spectral features, however, are strongly allied to the opacities. Thus the \Teff\ derived from the model atmospheres and theoretical evolutionary tracks are largely independent of each other. Our goal is to compare the \Teff\ inferred via the two methods, and draw lessons from the agreement or lack thereof.    

Most such tests so far have concentrated on field VLMS and BDs, and have uncovered discrepancies in \Teff\ between the atmospheric and evolutionary models \citep[e.g.,][]{Leggett01,Leggett02}. Very few studies, however, have carried out such model inter-comparisons for the PMS phase. \citet{Gorlova03} examined gravity indicators in model spectra compared to data for both field and PMS M-types; while they found various shortcomings in the synthetic atmospheres, they did not compare the evolutionary and atmospheric predictions against each other. \citet{Mohanty04a,Mohanty04b} did carry out such tests for a small sample of PMS VLMS and BDs, and found significant divergence between the two sets of models. However, the parameters they infer from model atmospheres are based on fitting narrow absorption features in high-resolution optical spectra, which does not address the overall accuracy of the continuum opacities in the synthetic spectra. 

In this paper, we take advantage of the large number of spectroscopically classified PMS VLMS and BDs for which near-infrared (NIR) to mid-infrared (MIR) photometry is now available (driven largely by {\it Spitzer} observations over the last decade). Most of the stellar flux for M types is emitted at these wavelengths, so the overall shape of the SED over this range is ideal for evaluating the broad continuum opacities in the synthetic spectra, and comparing the model atmosphere predictions to those of the evolutionary models. 

Our analysis proceeds as follows. We first derive the \Teff\ implied by the model atmospheres for PMS VLMS and BDs, by fitting the observed infrared (IR) SED (over 1.2--24\,$\mu$m) of young M-type spectral templates with synthetic photometry. We then compare these values to the \Teff\ predicted by the evolutionary models for these spectral types, given by the standard PMS spectral type-\Teff\ conversion scale. The details of how the two sets of predictions differ provide clues to the physical reasons for the differences. Finally, we use the atmospheric models to derive luminosities and temperatures for Class III (diskless) PMS VLMS and BDs of known spectral type in the Chamaeleon I (Cha I) and Taurus star-forming regions, and compare the results to theoretical HR diagrams. The outcome yields insights into the systematic errors in derived mass and age that arise from the discrepancies between the evolutionary and model atmosphere predictions.   

Our data are described in \S2, and the atmospheric and evolutionary models we test are discussed in \S3. We present our analysis of the M-type templates in \S4, and of the Cha I and Taurus sources in \S5. Various physical mechanisms to explain our results are detailed in \S6, and our conclusions summarized in \S7.  

\section{Sample Selection \& Templates}
Here we summarize the data used in our analysis. We describe our sample of Cha I and Taurus diskless PMS sources in \S2.1, the properties inferred in earlier studies for this sample (most importantly, the effective temperatures, based on evolutionary models) in \S2.2, and the PMS IR spectral type templates in \S2.3. 

\subsection{Selected Cha I and Taurus Class III Sources}
\subsubsection{Cha I}

Over 200 young sources have been classified as confirmed members of the Cha I star forming region, based on a suite of diagnostics including trigonometric parallax, luminosity classification, extinction, and spectral features characteristic of newborn stars \citep{Luhman04,Luhman07}. Spectral types have been assigned to almost all members, in the latter two papers, via comparisons of their molecular and atomic absorption features (TiO, CaH and VO bands and KI, NaI and CaII lines) to dwarf (for spectral types $\leq$M5) and averages of dwarf$+$giant (for types $>$M5) spectral templates. Some newer members have types determined through direct comparison to the spectra of previously classified members. 

Using {\it Spitzer} photometry, \citet{Luhman08} have investigated the disk properties of these sources based on their IR spectral slopes $\alpha$ (specifically, $\alpha_{K_s-24}$\footnote{Referred to as $\alpha_{2-24}$ by \citet{Luhman08}.} and $\alpha_{3.6-24}$). Of the 91 sources classified as Class III (i.e., diskless), 86 have $J H K_s$ photometry, either (in most cases) from 2MASS, or, in the absence of the latter, from ISPI\footnote{\citet{Luhman07} took NIR images of a dense subcluster in Cha I using the Infrared Side Port Imager (ISPI) at the 4\,m Blanco telescope at CTIO, to obtain NIR photometry of objects that do not appear in 2MASS.}. We remove two further objects because of large error flags in 2MASS and a lack of ISPI data, and exclude six more on the basis of being earlier than M in spectral type. Our final Cha I sample thus consists of 78 Class III M-type sources, all with NIR and MIR photometry and assigned spectral types (two of these have no available optical spectra, and have been assigned a rough spectral type of $\ge$M9 by \citet{Luhman07}, based on comparisons of their NIR spectra to those of other late-type objects). \citet{Luhman07} has also derived effective temperatures, extinctions and bolometric luminosities for these sources; these are described below in \S2.2. Our 78 Cha I sources, and their Luhman-derived properties, are listed in Table \ref{78objects}. 

\subsubsection{Taurus}
A large number of sources have also been verified as bona-fide PMS members of the Taurus star-forming region, and have had spectral types assigned, using the same membership criteria and spectral-type determination techniques described above for Cha I sources \citep[see][hereafter L2010, and references therein]{Luhman10}. With the aid of {\it Spitzer} photometry, and adopting the same IR spectral-slope ($\alpha$) diagnostics employed for Cha I objects, L2010 have further examined the disk properties of $\sim$99\% of the known Taurus members. Of the 352 sources listed in the latter paper, 119 are designated Class III (i.e., diskless). Further removing non-M types, as well as sources with no $\alpha_{K_s-8}$\footnote{Referred to as $\alpha_{2-8}$ by L2010.} data (used for deriving extinctions; discussed next and in \S2.2), we are finally left with 96 Class III M-type PMS objects in Taurus.     

Effective temperatures, extinctions, and bolometric luminosities have been derived for these sources in a series of earlier papers by Luhman and collaborators (see references in L2010), via the same methods employed for Cha I (see \S2.2). However, while L2010 use these parameters in various analyses, they do not cite their values explicitly; nevertheless, they do supply sufficient information to regenerate them. Hence, instead of trawling through the past papers, we simply rederive these quantities for our 96 sources from the data in L2010; comparing our inferred values to those cited in the original papers for a fraction of the sample, we find excellent agreement. Our methods are outlined in \S2.2 below. Our sample of Taurus M-type Class III sources, and their derived properties, are listed in Table 2. 

\subsection{Previously Derived Stellar Parameters for the Class III Objects} 
We describe here the techniques used by Luhman and collaborators to infer effective temperatures, extinctions and bolometric luminosities for the Cha I and Taurus sources. 

{\it In the rest of this paper, all parameters derived by such `Luhman' techniques are denoted by the subscript} `L': \LT\, for effective temperatures, \LAJ\, for ($J$-band) extinctions, and \LL\, for bolometric luminosities. 

\noindent {\it Effective Temperature}: The PMS spectral type-\Teff\ conversion scale currently used widely is founded on the seminal work of \citet{Luhman99}. Examining a large sample of PMS sources in the IC 348 star-forming region, Luhman showed first that the spectra of objects earlier than $\sim$M5 were well-matched by dwarf spectral templates, while later type PMS spectra were much better fit by averages of dwarf+giant templates. Furthermore, assigning the sources effective temperatures based on an extrapolation of the field M dwarf spectral type-\Teff\ scale inferred by \citet{Leggett96}, \citet{Luhman99} showed that mid-M and later sources in IC348 appeared significantly younger than earlier types when placed on a theoretical HR diagram based on the evolutionary models of \citet[hereafter BCAH98]{Baraffe98}; i.e., there seemed to be a systematic spectral-type-dependent non-coevality among the cluster members, which appears unphysical. Previous studies \citep{Luhman97,Luhman98} had noted this effect and, suspecting that it stemmed from problems in the adopted effective temperatures, explored using a spectral type-\Teff\ conversion scale for late-type PMS sources based on that of giants instead of dwarfs. \citet{Luhman99} argued that if the spectral features of low gravity mid-to-late-M PMS objects were intermediate between those of dwarfs and giants, then it was reasonable to advocate that their spectral type-\Teff\ conversion scale was intermediate between the two as well. He thus devised a new intermediate \Teff\ scale with the express purpose of imposing (mean) coevality on IC 348 members, as well as (precise) coevality on the components of the young quadruple system GG Tau in Taurus (since the members of a single system are very likely to be nearly coeval), {\it when using the BCAH98 evolutionary tracks}. The scale was later updated for M8/M9 PMS types by \citet{Luhman03}, using improvements to both cluster membership lists and dwarf temperature estimates, and including the evolutionary tracks of \citet[hereafter CBAH00]{Chabrier00}, which extend the BCAH98 models to lower masses (see \S3.1). This updated scale is the one we use for determining \LT\ for our PMS sources\footnote{Specifically, we use the \cite{Luhman03} scale for spectral types $\geq$M1; for $<$M1, we follow \cite{Luhman03} in using the dwarf temperature scale of \citet{Schmidt82}.}$^,$\footnote{\citet{Luhman07} provides a \Teff\ of 2350 K for an object of spectral type M9.25. As the \Teff\ scale supplied in \citet{Luhman03} ends at 2400K for M9, we linearly extrapolate from these two values to obtain 2200 K for L0.} (in particular, \citet{Luhman07} explicitly cites \LT\, for our Cha I sources based on this conversion scale; for our Taurus sources, we derive \LT\, using this scale and the spectral types provided in L2010). 

The crucial point here is that this PMS spectral type-\Teff\ conversion scale is explicitly constructed to agree with the BCAH98/CBAH00 evolutionary model predictions. In other words, {\it \LT\ may be regarded as the \Teff\ predicted by the BCAH98/CBAH00 theoretical evolutionary tracks for a given M spectral sub-type}. 

\noindent {\it Extinction}: Extinctions for individual Cha I sources are derived by \citet{Luhman07} by measuring the optical color excess over 0.6 to 0.9\,$\mu$m for each object at a given spectral type, relative to the bluest source at that spectral type (assumed to have zero extinction), and converting the excess into a $J$-band extinction \LAJ. For the few objects lacking optical data, \LAJ\, were derived using the \JH\ excess instead. 

For the Taurus sources, L2010 provide both the observed and dereddened values of the spectral slope $\alpha_{K_s-8}$, where the dereddening is accomplished using the \LAJ\ inferred (via the same techniques described above for Cha I) in previous papers. We thus rederive \LAJ\ for these objects by simply calculating the values needed to transform between the cited observed and dereddened $\alpha_{K_s-8}$ (using $A_J/A_{K_s} = 2.5$ from \citet{Indebetouw05} and $A_{[8.0\,\mu{\mathrm m}]}/A_{K_s} \approx 0.49$ from \citet{Flaherty07}: same ratios as used by L2010). 

\noindent {\it Luminosity}: Finally, bolometric luminosities \LL\ for Cha I sources were derived by \citet{Luhman07} by applying bolometric corrections from \citet{Kenyon95} \& \citet{Dahn02} to the dereddened $J$-band photometry, and using a mean distance modulus of 6.05 (corresponding to $d \approx$ 162\,pc) for Cha I. 

For the Taurus sources, we first deredden the observed $J$-band photometry supplied in L2010 using the \LAJ\, calculated above, and then derive \LL\ by applying the same bolometric corrections used for Cha I, employing a mean distance modulus of 5.73 (corresponding to $d \approx 140$\,pc).

\subsection{PMS Spectral Type Templates}
Intrinsic IR colors of PMS stellar photospheres as a function of spectral type have been collated by L2010, spanning the range K4 to L0\footnote{The [$K_s$$-$3.6] colors provided for PMS spectral types M4--M9 in Table 13 of the original L2010 paper were erroneous; these were corrected in the erratum subsequently published for that paper, and we have used the corrected values.}. These spectral type-color `templates' were derived by fitting an envelope to the bluest IR colors of young objects as a function of spectral type, the rationale being that the bluest sources are likely to be those least affected by dust extinction and excess disk emission, and thus representative of the naked photospheric colors of PMS sources. The young objects used to derive the templates were selected from the Cha I and Taurus star forming regions and the $\eta$ Cha, $\epsilon$ Cha, TW Hya associations, supplemented by a few young solar neighbourhood stars at the latest types. 

\section{Models}

\subsection{Theoretical Evolutionary Tracks}
We employ the theoretical evolutionary tracks by BCAH98 and CBAH00, on which the current PMS spectral type-\Teff\ conversion scale is based (as described in \S2.2). As they encompass different mass ranges (BCAH98 span 0.02-1.4\Msun\ while CBAH00 span 0.001-0.1\Msun), we incorporate them over different limits. The main difference between the two is in the treatment of dust: CBAH00 include grain opacity for the lower masses (for which dust becomes important), while BCAH98 neglect it entirely. As noted in \S1, the temperature and luminosity evolution is predominantly controlled by interior conditions, and only weakly dependent on the atmospheric opacities, so this difference in grain treatment is not of any great consequence in our investigation of these quantities. Indeed, for the modest range in mass over which the two tracks overlap, the predicted PMS stellar temperatures and luminosities (and hence also radii) are nearly identical. We therefore combine the tracks by adopting BCAH98 for $M_*\ge0.02$\Msun, and CBAH00 for lower masses. These evolutionary models measure the efficiency of convection in the stellar interior using the mixing length parameter\footnote{$\alpha = l_{mix}/H_P$, where $l_{mix}$ is the mixing length and $H_P$ is the pressure scale-height.}, $\alpha$. Initially, BCAH98 offered a choice of $\alpha=1.0$, 1.5, or $1.9$ (where $\alpha=1.9$ is required to fit the sun) above 0.6\Msun, whilst below this mass they gave only $\alpha=1.0$. CBAH00 also provide only $\alpha=1.0$ for the lowest masses. This is because at main sequence ages, masses $\gtrsim 0.6$\Msun\ ($\lesssim$\,M0) have extended superadiabatic layers, making their evolutionary modelling sensitive to the precise choice of $l_{mix}$; for lower masses, these layers have receded, such that the exact choice of $l_{mix}$ is no longer important \citep{Baraffe98}. However, it soon became apparent that at the low surface gravities found at young ages, these extended superadiabatic layers (and thus the choice of $l_{mix}$) begin to affect lower masses \citep{Baraffe02}. The BCAH98 tracks for $\alpha=1.9$ were then extended down to 0.1\Msun, to incoporate the full mass range at which the choice of $l_{mix}$ is important. Indeed, comparing the tracks for $\alpha=1.0$ and 1.9 by eye it is clear that, at young ages ($\tau \sim 1$\,--\,5\,Myr), small deviations between the two start to appear at masses $\gtrsim 0.3$\Msun\ ($\lesssim$\,M4), becoming prominant by $\gtrsim 0.4$\Msun\ ($\lesssim$\,M3). As such, we have used the BCAH98 models with $\alpha=1.9$ for\footnote{Though the tracks with $\alpha=1.9$ were not available below 0.6\Msun\ at the time, \citet{Luhman03} noted that young objects with dynamical masses above this limit favoured the $\alpha=1.9$ models over $\alpha=1.0$, suggesting it was the better fit when $l_{mix}$ does become important.} $M_*\geq 0.1$\Msun, and $\alpha=1.0$ for $M_*< 0.1$\Msun.

\subsection{Atmospheric Models}
We use the latest version of synthetic spectra generated with the PHOENIX code, namely the AMES-Cond/AMES-Dusty \citep{Allard01} and the BT-Settl \citep{Allard12a,Allard12b} models. The AMES models, with now outdated solar abundances and line lists, have for many years been widely used in modelling young stars and are closely associated with the BCAH98/CBAH00 evolutionary tracks. The BT-Settl model uses updated solar abundances which reduces the amount of oxygen, an important species in cool objects (appearing in TiO, H$_2$O, VO and CO throughout the optical/IR); in this work we use the BT-Settl model with the \citet{Asplund09} abundances. Improvements of the line lists in the previous decade have also been included; for a full discussion see \citet{Allard12b}. Due in part to these advances in our knowledge of atmospheric composition, the BT-Settl models have been recently shown to roughly fit the observed NIR colors in the main sequence from spectral types M to T, something which has eluded previous generations of models \citep{Allard12b}. 

The primary difference between the atmospheres of these three models stem from their assumptions of photospheric dust; dust grains are allowed to form in each, but differ in where they end up after formation. Negligible dust forms in the models above $\sim 2600$ K and we therefore expect all models to behave similarly in this region \citep{Allard01}. Below this temperature, differences between the models should become apparent due to how they treat the onset of dust.

\noindent {\it AMES-Dusty}: In these models, dust grains are incorporated in both the chemical equilibrium and opacity calculations. The aim here is to neglect gravitational settling and allow the build up of grains in the photosphere, which observationally happens over the late M to late L types.

\noindent {\it AMES-Cond}: In these models, dust grains are present in the chemical equilibrium calculations but not in the opacity lists. This imitates the grains forming and then immediately gravitationally settling out of the atmosphere. Observationally, this appears to happen in objects below 1500 K (the T spectral class), which are too cool to allow grains to remain suspended in the atmosphere.

\noindent {\it BT-Settl}: In these models, the aim is to go one step further and mimic the true effect of dust grains by allowing them to gradually settle out of the atmosphere as the temperature decreases, emulating the observed transition in color from red to blue at the L-T transition. Note that at different temperature limits, the BT-Settl model should behave like either the AMES-Cond and AMES-Dusty; above $2600$ K, all three models should be similar with any differences mainly due to line lists/abundances. Just below $2600$ K, whilst still relatively warm, grains have formed and remain suspended in the photosphere. In this scenario the BT-Settl models should resemble the AMES-Dusty models. As the temperature decreases further, the dust grains will start to condense down from the photosphere, contributing less and less to the atmospheric opacities. Beyond the point at which no grains are left in the atmosphere ($<1500$ K), the BT-Settl models will act similar to AMES-Cond.

We will also briefly consider another limiting case of cloud formation, the {\it NextGen} model \citep{Allard97,Hauschildt99}, in which no dust forms at all.

According to theoretical evolutionary tracks, PMS objects should still be quite large and thus have lower surface gravities than main sequence dwarfs. For M type objects at the age of Cha I this corresponds to \logg \ between $3.5$ and $4.0$. We will focus mainly on synthetic spectra with \logg \ $=4.0$ in this analysis, as justified later. \citet{Padgett96} analysed a number of young, nearby star forming regions, including Chamaeleon and Taurus, and found roughly solar metallicity abundances ($\left|[Fe/H]\right|<0.1$) in each of them; therefore we will also only use synthetic spectra at solar metallicity. 

\section{Results I - Comparisons between Synthetic and Template Photometry}
On the Main Sequence, the merits of the various atmospheric models discussed above are often evaluated based on comparisons to the \JK\ colors of dwarf spectral templates \citep[e.g.,][]{Allard12a,Allard12b}. We therefore first reproduce this exercise for the PMS case, by comparing the models to the \JK\ colors of the L2010 PMS spectral templates (\S4.1). This allows us to identify general trends in the models, and assess their overall compatibility with the data. We then undertake a detailed multi-band comparison between model and template photometry over the full wavelength range available for the templates, spanning 1.2 to 24\,$\mu$m (\S4.2). This enables us to to define a new PMS spectral type-\Teff\ conversion scale based on the synthetic atmospheres, and compare it to the independent temperature scale based on evolutionary models.               

\subsection{$J$-$K_s$ Colors}

\subsubsection{General Trends in Synthetic $J$-$K_s$}
Fig.\,1 shows the synthetic and template \JK\ colors as a function of \Teff. We first discuss the trends in the models (curves in Fig.\,1). 

All the atmospheric models are in very good agreement above a model-\Teff\ of $\sim$3900\,K, where the main sources of NIR opacity are H$^-$ and H$_2^-$. At lower temperatures, H$_2$O formation becomes increasingly important (with the molecule dominating the NIR opacity by $\sim$3300\,K: \citealt{Allard94, Allard95}, and thus the difference in elemental abundances between AMES-Dusty/Cond on the one hand and BT-Settl on the other (in particular, the lower oxygen fraction, and hence depressed H$_2$O formation, in the latter) causes BT-Settl to deviate redwards of Dusty/Cond in this regime. The NextGen model, with much older abundance and opacity lists, appears even redder. At \Teff\,$\lesssim$ 2600\,K \citep{Allard12b}, dust formation becomes efficient in the synthetic atmospheres. Atmospheric grains act as a strong opacity source in the optical and IR, and, by destroying H$_2$O via backwarming, simultaneously reduce H$_2$O opacity in the IR; the combined effect is to suppress the flux emitted at shorter wavelengths and enhance it at longer, making the overall spectrum redder than in the absence of dust. Consequently, the Dusty model (where grains remain suspended in the photosphere) becomes much redder at these temperatures than Cond (where grain opacity is neglected). Conversely, Dusty and BT-Settl converge by 2300\,K and remain very similar down to 1800\,K, since dust opacity dominates in this regime for both models (i.e., the gradual settling of grains in BT-Settl does not have an appreciable effect until still lower \Teff, not shown). Finally, without any dust formation at all, the NextGen model cannot keep apace of the rapid reddening in \JK\ with decreasing \Teff\ in Dusty and BT-Settl, and eventually becomes bluer than the latter models. 

Fig.\,1 also shows the behaviour of the Dusty model at \logg\, = 3.5 versus 4.0 (which captures the range in surface gravities expected from evolutionary models at PMS ages for low mass stars and brown dwarfs). We see that the two are nearly indistinguishable; the same is true of Cond and BT-Settl models over this range of gravities as well (not shown). Indeed, the model trends discussed above are very similar to those observed in the same models at the significantly higher gravities (\logg\, $\approx$ 5.0--5.5) appropriate to 
Main Sequence field dwarfs \citep[see Fig.\,3 in][]{Allard12b}. In other words, the basic opacity behaviour that leads to these trends appears quite insensitive to the surface gravity in the synthetic atmospheres.  
 
\subsubsection{Comparisons with Template $J$-$K_s$} 
Next, we compare the synthetic \JK\ colors to those of the spectral templates. Fig.\,1 shows \JK\ for the L2010 PMS templates plotted as a function of the Luhman effective temperature scale \LT, where the latter is based on the evolutionary models (see \S2.3). A few systematic trends with spectral type are clear.   

First, spectral types M0 and earlier are slightly bluer than all the models, by $\sim$0.075\,mag. From M1 to M6, the templates are in fairly good agreement with AMES-Cond/Dusty and BT-Settl, with Cond/Dusty becoming slightly bluer than the templates, and BT-Settl providing a better fit, over M4--M6. Finally, M7 and later templates drift increasingly redward of the Dusty and BT-Settl models (and even more so compared to Cond), with a deviation of $\sim$0.65\,mag relative to Dusty/BT-Settl by L0. The insensitivity of the synthetic spectra to surface gravity, discussed above, means that these trends are independent of the precise \logg.   

We also compare the templates to the NextGen model, even though the latter is based on quite outdated opacities and abundances, in order to make an important point. In general, NextGen is a poor match to the templates, being significantly redder than the \JK\ data for spectral types earlier than M8 and much bluer at M9 and later. It does, however, provide a good match at M8. This does not mean, though, that NextGen reproduces the shape of the spectrum from $J$ to $K_s$ at M8: closer inspection (not plotted) reveals that the model \JH\ is much redder than in the template, while the model \HK\ is much bluer, with the opposing offsets coincidentally cancelling out to give a perfect match to \JK\ at M8. This implies that it is not sufficient to examine a single color, spanning multiple photometric bands, to gauge the merit of the atmospheric models; it is necessary instead to carry out a detailed comparison over all the available template photometry. We embark upon this in the next section (4.2), and show that the agreement in \JK\ mentioned above, between the mid-M PMS templates and BT-Settl, is also largely illusory, arising, as in the NextGen case above, from the cancellation of opposing deviations in \JH\ and \HK.   
 
These trends are very similar to those found for field dwarfs \citep[][Fig.\,3]{Allard12b}. BT-Settl, incorporating the gradual settling of grains, overall provides the best fit to \JK\ in dwarfs spanning the entire range M to T; however, the same deviations we find for PMS templates at the K/M transition and at late M types are apparent in the field as well, with the dwarf templates being slightly bluer than BT-Settl in the former regime and redder in the latter. \citet{Allard12b} suggest that the discrepancy at the K/M boundary may arise from an under-representation of K dwarfs in the data, while the deviation at the late Ms may be due to a missing ingredient in model dust opacities, such as large porous grains. In \S6.3, we propose a more physical explanation, at least for the PMS case, for the offset at the K/M transition, and also suggest a general underestimation of dust opacities to explain the deviation in young late M objects. Finally, it is not clear whether the good match in \JK\ between BT-Settl and the early-to-mid-M field dwarfs is real, or due to opposing trends in \JH\ and \HK\ as in our PMS case; this needs to be investigated in future studies.  

\subsection{Model-Fitting Over 1.2 to 24\,$\mu$m} 
We now perform a statistical analysis to find the best fitting synthetic spectrum from the atmospheric models for each spectral type template, using all 7 available template IR colors (from 8 IR photometric bands: $J$, $H$ and $K_s$ from 2MASS, and 3.6\,$\mu$m, 4.5\,$\mu$m, 5.8\,$\mu$m, 8.0\,$\mu$m and 24\,$\mu$m from {\it Spitzer}). 

We first interpolate between the templates to construct the same 0.25 spectral-subclass grid spacing for the template colors as supplied for the Cha I/Taurus sources (since we eventually wish to assign model atmosphere temperatures to the individual objects in these star-forming regions; see \S5). 

In fitting model atmospheres to the templates, we have a choice: we can either compare the two on the basis of colors (by first computing the model fluxes in each photometric band, then deriving the implied synthetic colors, and finally comparing to the template colors supplied), or on the basis of fluxes in individual photometric bands. The latter is statistically more robust: in the former case, errors across adjacent colors are correlated, vitiating the statistical interpretation of the quality of the fits. To convert the template colors to photometry, we first set the template $J$-band flux to an arbitrary initial value (say, 0 mags), and use the known colors to calculate the fluxes in the remaining bands relative to this value. Errors in the template photometry are not supplied by L2010; consequently, we assign mean errors to each of the 8 IR photometric bands based on the average 2MASS and {\it Spitzer} photometric errors cited for the Cha I and Taurus sources in \cite{Luhman07} and \cite{Luhman08} (which comprise the main sample used to derive the L2010 spectral templates). The resulting template errors are $\sim$0.025\,mag in the 2MASS $J$$H$$K_s$ and {\it Spitzer} IRAC 3.6\,$\mu$m, 4.5\,$\mu$m bands, $\sim$0.035\,mag in the {\it Spitzer} IRAC 5.8\,$\mu$m, 8.0\,$\mu$m bands, and $\sim$0.045\,mag in the {\it Spitzer} MIPS 24\,$\mu$m band. 

Next, synthetic photometry is derived from the model spectra. The latter are supplied at intervals of 100\,K in \Teff; we interpolate between these to construct a model grid-spacing of 25\,K to improve the precision of fits to the spectral templates. The spectra are then convolved with the appropriate filter transmission curves\footnote{The 2MASS bandpasses were taken from \citet{Cohen03}, IRAC from \citet{Hora08}, and the MIPS (24\um) bandpass from the NASA/IPAC website: \url{http://irsa.ipac.caltech.edu/data/SPITZER/docs/mips/calibrationfiles/spectralresponse/}} to derive the model photometry in each of the 8 IR bands, using the spectrum of Vega \citep{Cohen92} for calibration, following the method outlined in \citet{Buser92}\footnote{The resolution of the AMES-Dusty spectra decrease rapidly after 10\um; we therefore interpolate over wavelengths in this regime to achieve the same resolution as in AMES-Cond, to improve the accuracy of our 24\,$\mu$m photometry.}. We note that these values represent the model photometry {\it at the stellar surface}; i.e., they are independent of the stellar radius and distance (this fact is used later to calculate stellar radii and luminosities for individual sources; see \S5.2). Errors in the synthetic photometry, arising from uncertainties in the computation of the model spectra, are negligible compared to the observational errors in 2MASS and {\it Spitzer} photometry, and thus ignored. 

Finally, the best fit between a given spectral template and each model atmosphere is found by scaling the template photometry till the 
root mean square (rms) deviation between the template and model values is minimised. The rms deviation is defined here as
\be
\sigma = \sqrt{\frac{1}{N}\sum \limits_{i=1}^{N}\left(\frac{x_i-\mu_i}{\sigma_i}\right)^2}
\label{RMS}
\ee
where $N$ is the total number of bands (=8), $x_i$ the template photometry in the $i$-th band (including some scaling factor), $\mu_i$ the corresponding model photometry, and $\sigma_i$ the observed photometric error in that band. The best-fit model atmosphere overall is the one that yields a global minimum in $\sigma$ over the entire range of model \Teff\ tested (in our case, over 1300--4300\,K, comfortably bracketing the plausible range in \Teff\ for L0--M0 PMS sources).{\it We term this best-fit temperature \AT\, (denoting \Teff\ from `Allard' atmospheric models)}. 

In addition, for any given spectral-type template, we also examine the goodness-of-fit to the model atmosphere corresponding to the evolutionary track-predicted temperature (\LT) for that spectral type, to quantitatively probe the difference between the atmospheric and evolutionary predictions. 

\begin{figure}[h!]
\begin{center}
\includegraphics[width=17cm]{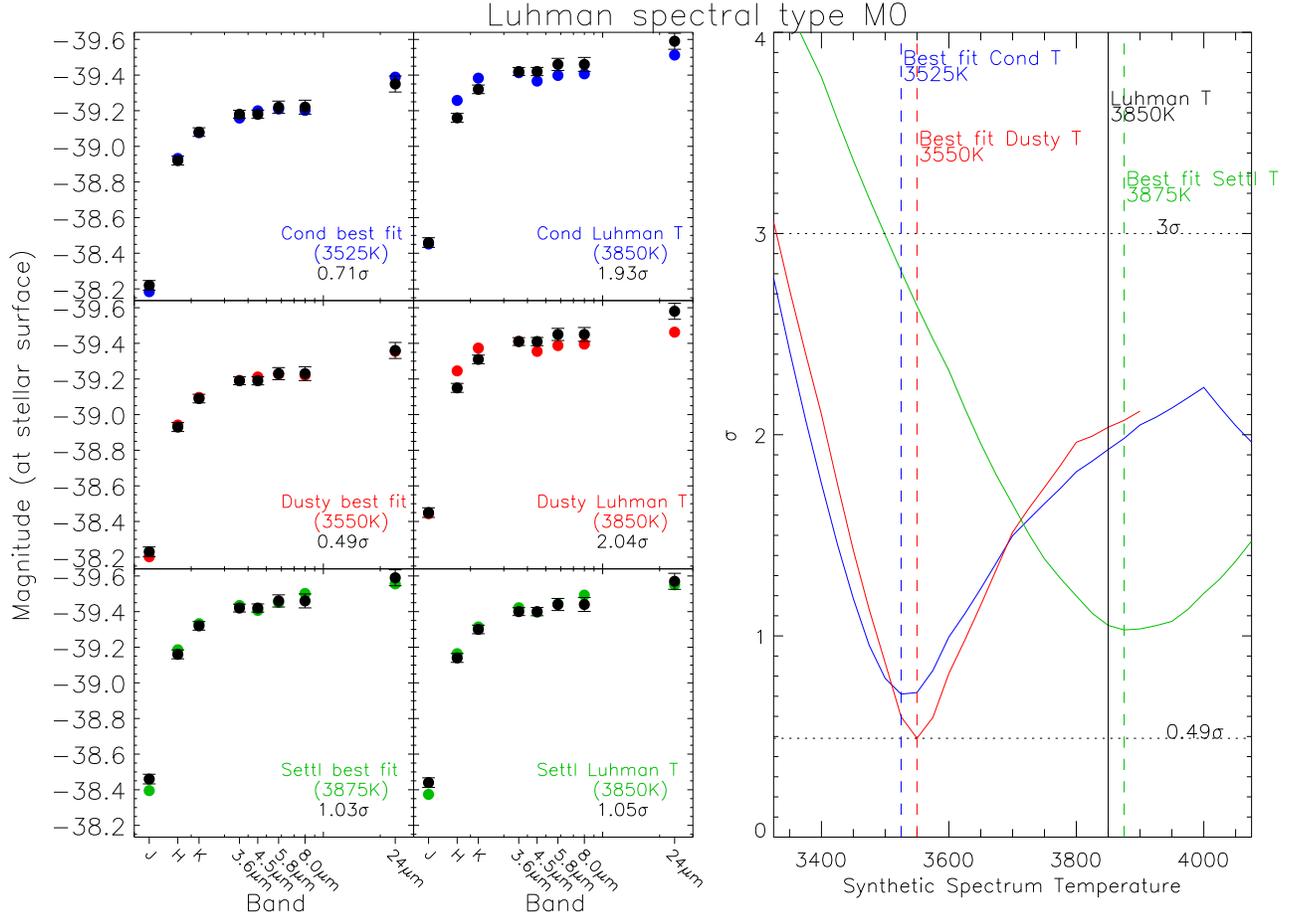}
\caption{A goodness of fit for the three models considered in this paper, AMES-Cond (blue), AMES-Dusty (red), and BT-Settl (green), all at \logg $=4.0$, against photometry derived from the M0 spectral color template given in L2010. Left panels: the fits at both \LT \ (to $\pm$25K) and the best fit temperature for each model. Temperature and rms value are indicated in the bottom right of each subplot. Right panel: The rms curve across a reasonable temperature range. The rms best fit and an arbitrary 3$\sigma$ line are indicated on the plot, as are the Luhman \& best fitting temperatures for each model.}
\label{M0}
\end{center}
\end{figure}

\begin{figure}[h!]
\begin{center}
\includegraphics[width=17cm]{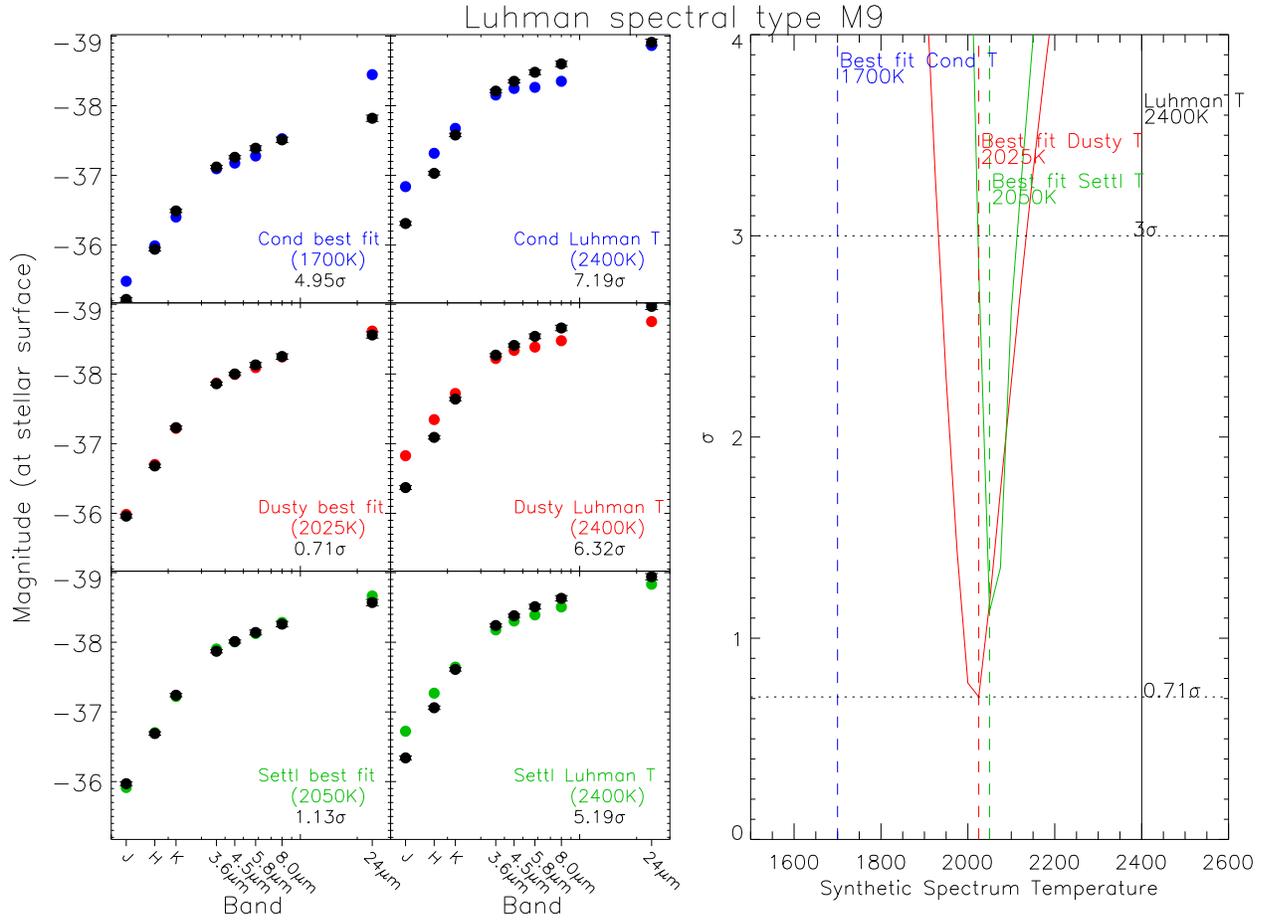}
\caption{Same as Fig.\,\ref{M0}, for M9. Note the AMES-Cond fits don't make an appearance on the right panel, due to its best fit being worse than 4$\sigma$.}
\label{M9}
\end{center}
\end{figure}

Our results are tabulated in Table \ref{LTvsAT}, shown in part in Figs.\,\ref{M0} - \ref{M9}, and summarised in \S4.2.1-4.2.3 below for early, mid, and late M types respectively. Before proceeding to the latter discussion, we go over the first of these plots (Fig.\,\ref{M0}, for spectral type M0) in some detail for explanatory purposes. In the right hand panel, we plot the standard deviation between the template and each model atmosphere against the model \Teff, with the red, blue and green curves denoting AMES-Dusty, AMES-Cond and BT-Settl models respectively. The $\sigma$ value of the best fit, as well as the $3\sigma$ limit, are shown as horizontal lines. The vertical lines mark the best-fit temperature \AT\, (i.e., the model \Teff\ with the lowest global $\sigma$) for each model, as well as the value of \LT\, (the evolutionary model-predicted \Teff) for the spectral type under consideration (M0 in this case). Note that the curves for AMES-Dusty and BT-Settl stop at 3900 K and 4000 K respectively, as these are the highest \Teff\,available for these models (this does not affect our results, since the best-fits are obtained at comfortably lower temperatures). The 6 plots in the left panel show the fits to the data for each model, both at the best-fit temperature \AT\, for that model as well as at the evolutionary track-predicted temperature \LT\ for this spectral type. We see that for M0, AMES-Dusty provides the best fit, but at a \AT\ several hundred Kevlin lower than the evolutionary track expectation \LT. The best fit to BT-Settl, on the other hand, is at a \AT\ similar to \LT, but the overall quality of this fit is worse than for the best fit to AMES-Dusty, due to deviations at $J$ and 8.0$\,\mu$m. To investigate the trends across the entire M spectral range, we split our templates into three groups: early M (M0-M2), mid-M (M3-M6) and late M ($\geq$ M7, and including L0).

\subsubsection{Early M (M0--M2)}
In the early M types (e.g., M0, Fig.\,\ref{M0}), the AMES-Cond and AMES-Dusty models are both somewhat bluer than the observed IR SEDs at the temperatures \LT\ predicted by the evolutionary models for these spectral types. The deviation is largest at M0, and declines with later type. Reducing the effective temperature makes the models redder; consequently, the best fits to these spectral types are obtained at temperatures \AT\, lower than \LT: by 300\,K at M0, and 100\,K at M2. Note that the AMES-Cond and AMES-Dusty best-fits are nearly indistinguishable, which is not surprising given that the latter models differ primarily in their treatment of dust, and grains do not form at all in the synthetic atmospheres at these \Teff.   

Conversely, the BT-Settl models are slightly redder than the observed SEDs at the evolutionary model-predicted temperatures \LT, with the deviation increasing from M0 to M2. The best fits to these models are thus obtained at \AT\, somewhat higher than \LT: by $\sim$25\,K at M0, and 100\,K at M2. Note that the quality of the best fits to BT-Settl is always worse at these spectral types than that of the best fits to AMES-Dusty. On the other hand, at \LT\, itself, the fits to BT-Settl are superior than to AMES-Dusty (or AMES-Cond), with nearly all the deviation from the data appearing at only $J$ and 8.0\,$\mu$m. We return to this point in \S6.3.  

\subsubsection{Mid M (M3--M6)}
\begin{figure}[h!]
\begin{center}
\includegraphics[width=17cm]{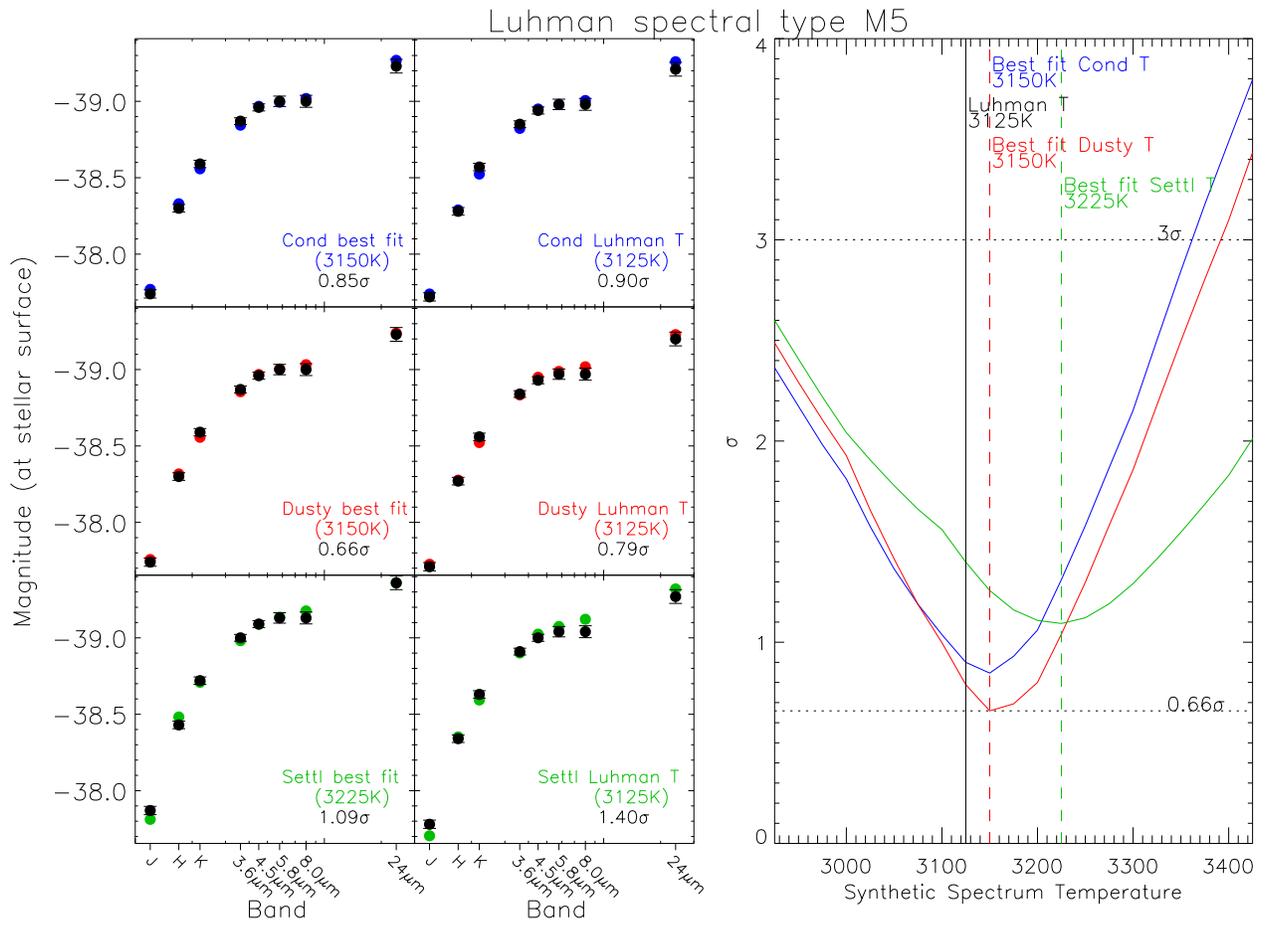}
\caption{Same as Fig.\,\ref{M0}, for M5}
\label{M5}
\end{center}
\end{figure}

For the mid-M types (e.g., M5, Fig.\,\ref{M5}), the AMES-Dusty and AMES-Cond IR SEDs at \LT\ are very similar to the observed ones, and the best fits to these models are obtained at \AT\, differing by just $\leq$100\,K from \LT. 

\begin{figure}[h!]
\begin{center}
\includegraphics[width=10cm]{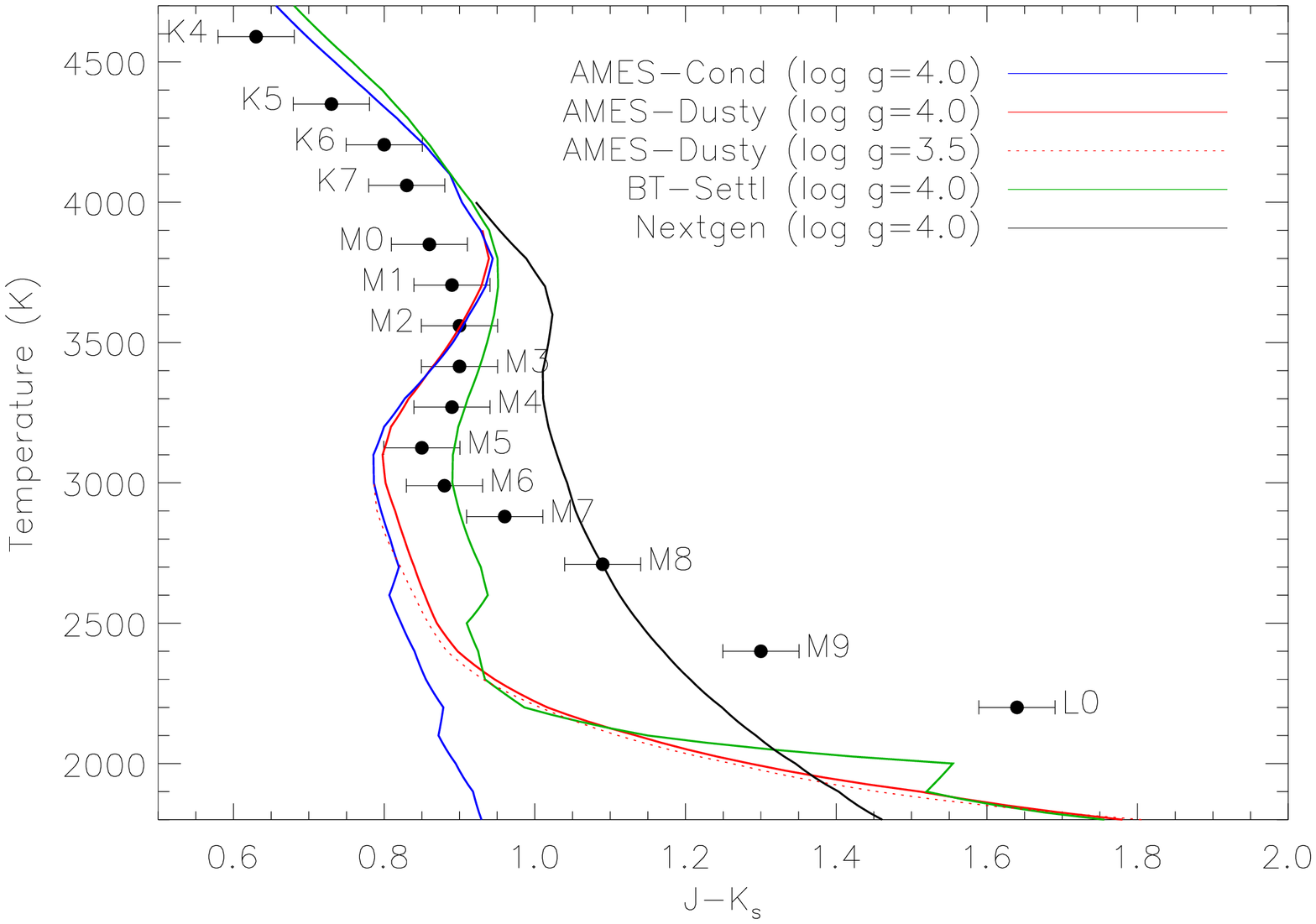}
\caption{Model \JK \ colors (model version shown in the legend) compared with the L2010 PMS spectral template colors (converted using the PMS spectral type-\Teff \ (\LT) scale from \citet{Luhman03}). This is analogous to Figs.\,3 \& 6 in \citet{Allard12b}, but uses PMS objects instead of Main Sequence.}
\label{allard12comparison}
\end{center}
\end{figure}

The best fits to BT-Settl at these spectral types, on the other hand, are generally obtained at \AT\, differing by $\gtrsim$100\,K from \LT; moreover, the quality of the best fits to these models is inferior to those obtained for AMES-Dusty. This seems at odds with the excellent match in \JK\ color at \LT\ between BT-Settl and the mid-M types, plotted in Fig.\,\ref{allard12comparison}. A closer perusal of Fig.\,\ref{M5} (bottom right plot in left panels) reveals the reason: at \LT, \JH\ in these models is redder than the data, while \HK\ is bluer. The two effects cancel to give a nearly perfect match between BT-Settl and the observed \JK\ at these spectral types in Fig.\,\ref{allard12comparison}, but the detailed shape of the model SED from $J$ to $K_s$ remains a poor fit to the data. As advertised in \S4.1.2, this mirrors the spuriously good \JK\ fits to the outdated NextGen models at certain spectral types, and underlines the need for multi-band analysis to establish the veracity of the synthetic atmospheres. 

\subsubsection{Late M ($>$M6)}
In the late M types (e.g., M9, Fig.\,\ref{M9}), all the atmospheric models once again become bluer than the observed IR SEDs at the evolutionary model-predicted temperatures \LT, with the deviation increasing with later type. As a result, the best synthetic atmosphere-fits to the data are obtained at \AT\, significantly lower than \LT, by $\sim$200--500\,K. AMES-Dusty and BT-Settl perform similarly in this regime, though formally the best fits to AMES-Dusty are generally superior, with BT-Settl yielding better quality best-fits only around M7 and L0. At spectral types M8 and later, the best fits to AMES-Cond occur at much lower \Teff, and are of much worse quality, than those obtained for AMES-Dusty/BT-Settl; this is because dust formation becomes important in the best-fit synthetic atmospheres at these spectral types, and it is the dust opacity that allows the very good fits to Dusty/Settl here, while dust opacity is neglected in the Cond models. Finally, note that in the latest types, around M9/L0, the shape of the model SED varies rapidly with \Teff\ (due to accelerating grain formation), leading to a very narrow range of \Teff\ over which good fits to Dusty/Settl can be attained. 

\begin{figure}[h!]
\begin{center}
\includegraphics[width=10cm]{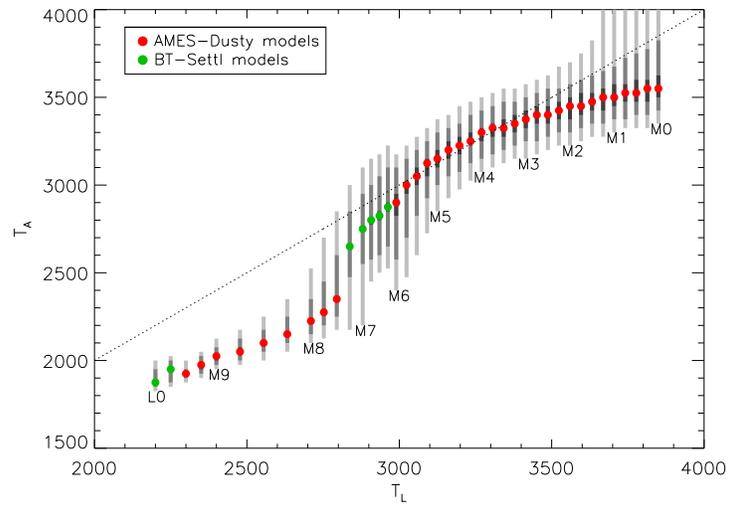}
\caption{\AT \ vs. \LT, showing how the synthetic spectra predict lower temperatures in both the early and late M types than derived by Luhman. The grey bars, in order of decreasing strength, highlight the range in temperatures which give better fits than 1$\sigma$, 2$\sigma$, and 3$\sigma$, respectively. The type of model that provides the best fit, in this case either AMES-Dusty (red), or BT-Settl (green) is also indicated.}
\label{2sig}
\end{center}
\end{figure}

In conclusion: {\it (1)} AMES-Dusty atmospheric models generally out-perform BT-Settl (and AMES-Cond) in providing better fits to the template IR SEDs across the entire M spectral type range. In the few cases where BT-Settl yields a better quality best-fit (around M7 and L0), the implied atmospheric \Teff\ remains very similar to that obtained from AMES-Dusty. {\it (2)} For spectral types earlier than M8, the synthetic IR SEDs vary slowly with \Teff, producing a relatively broad range of \Teff\ over which acceptable fits to the data are obtained (defined here as within a 3$\sigma$ rms deviation from the template SED, with $\sigma$ defined by Equation (1)). At M8 and later types, the synthetic SEDs vary rapidly with \Teff, yielding a very narrow range of acceptable temperatures. {\it (3)} In both the early and late M types, the best-fits to the model atmospheres are obtained at \AT\, significantly lower than the evolutionary model predictions \LT\ for these types, by $\sim$300\,K in the early Ms and $\sim$500\,K in the late Ms. It is only in the mid-M types that \AT\ and \LT\ converge. These conclusions are illustrated graphically in Fig.\,\ref{2sig}, and tabulated in Table \ref{LTvsAT}.  

\section{Results II - Comparisons to Cha I and Taurus Sample}
We have now assigned a best-fitting synthetic spectrum to each M spectral subclass, based on empirical IR template colors. The corresponding model temperatures \AT\ thus define a new spectral type-\Teff\ conversion scale. We can use this to reassign temperatures -- based on synthetic spectra alone, independent of evolutionary models -- to our individual Cha I and Taurus sources with known spectral type. Moreover, we can also derive luminosities and radii for these sources using the atmospheric models, again independent of evolutionary predictions. Doing so allows us to compare the predictions of the evolutionary models and synthetic atmospheres for various stellar parameters. We first summarize our methods for calculating stellar luminosities and radii using the model atmospheres, in \S5.1, and then discuss our results in \S5.2.  

In keeping with our nomenclature so far, {\it quantities inferred using the atmospheric models are denoted below by the subscript `A' (for `Allard' atmospheres)}: $BC_{J,A}$ for ($J$-band) synthetic bolometric corrections, \AL\ for bolometric luminosities and $R_A$ for stellar radii.  

\subsection{Deriving Luminosities and Radii from Model Atmospheres}
\subsubsection{NIR Extinctions}
Before going on to derive luminosities and radii from the model atmospheres, we must revisit the extinctions for our Cha I and Taurus sources. As discussed in section 2, extinctions have already been inferred for our sample by Luhman and collaborators, based for the most part on optical spectra. However, our foregoing temperature analysis is based on {\it infrared} spectral templates. To minimise errors arising from mismatches between the optical and infrared, we rederive the extinctions for our sources directly in the infrared, by comparing their observed NIR photometry to those of the L2010 spectral templates. 

For a given Cha I/Taurus source of known spectral type, we compute the NIR extinction, \AAJ, by comparing its $J$$H$$K_s$ photometry to that of a template of the same spectral type\footnote{The spectral types of our sources have been determined by comparison to the spectral templates as well; however, the latter determinations are based on individual absorption features spanning narrow wavelength regimes over which the extinction remains very nearly constant, and so does not influence the inferred spectral type. In other words, our extinction determination, based on a broad-band comparison to the templates, is not vitiated by the use of the same spectral templates for spectral typing.}, and employing the reddening law from \citet{Indebetouw05}\footnote{$\frac{A_J}{A_{K_s}}=2.5$ and $\frac{A_H}{A_{K_s}}=1.55$.}. The method used to find the best-fit \AAJ\ -- minimising the rms deviation between the source and template photometry -- is exactly the same as described in \S4.2 for determining the best-fit synthetic SED to a spectral template, except the synthetic SED is now replaced by the SEDs of a particular Cha I and Taurus sources, and the comparison is only over the NIR $J$$H$$K_s$ bands (so $N=3$ in equation (1)). During fitting, extinctions are also allowed to be (unphysically) negative, to flag any pathological sources; objects with a negative best-fit \AAJ\ are dealt with individually (see below). 

Our \AAJ\ are compared to the previously derived \LAJ\ in Figs.\,6(a) and (b), for Cha I and Taurus sources respectively. The two extinction estimates are in good agreement, with no systematic offsets; the difference between \AAJ\ and \LAJ\ is mostly $\lesssim$\,0.2\,mag (\citet{Luhman07} states an error of $\sim$\,$\pm$0.13 mag for his \LAJ\ estimates; the somewhat larger deviations observed here between \AAJ\ and \LAJ\ are expected to arise from the difference in wavelength regime -- optical versus NIR -- used to calculate the two). Our derived \AAJ\ for the Cha I and Taurus objects are listed in Tables 1 and 2 respectively. Objects with a large negative best-fit \AAJ\ (1 source in Cha I; see below), large ($>$\,0.5\,mag) difference between \AAJ \ and \LAJ \ (3 in Cha I, 2 in Taurus), unreasonably poor best-fits ($\sigma > 1$; 8 each in Cha I and Taurus), or some combination of these, are flagged in Tables 1 and 2, as well as in Figs.\,8 and 9.

Only one object in Cha I, 2M J1110-7642, had a negative best-fit \AAJ\ ($-$0.38 mags). We set its extinction to zero (which is also its \LAJ\ from \citet{Luhman07}) for the remainder of this analysis, but flag it as a poor extinction fit in Table 1 and Figs.\,8 and 9. In Taurus, we find negative best-fit extinctions for 10 objects. However, these are all still quite close to zero, with even the most extreme case being \AAJ \ = $-$0.18 mag; moreover, none of these differ from \LAJ\ by more than 0.5 mag. Consequently, \AAJ\ for these sources are set by fiat to 0.0 mag (and are not flagged in Table 2 or in the plots).

\subsubsection{Luminosities and Radii}
With more precise IR extinctions in hand, as well as \AT\ from the model atmospheres, we compute the radii and luminosities implied by the latter models for our sample. Our method essentially amounts to applying a synthetic bolometric correction based on the atmospheric models. Since the empirical bolometric corrections used to infer \LL\ for our sources are in the $J$-band, we perform our analysis at $J$ as well.  

In what follows, we assume that the best-fit synthetic SED we have found to every spectral type template is a {\it perfect} fit; i.e., we ignore any remaining deviations at $J$ between the template and its best-fit model ($\lesssim$\,0.05\,mag in all cases; see Figs.\,2--4). Now recall, from \S4.2, that the synthetic fluxes refer to the emission at the stellar surface. Consequently, the dereddened observed $J$-band flux of any source at a given spectral type, denoted by $F_{J, source}$, is related to the corresponding model $J$-band flux (at the \AT\ for that spectral type), $F_{J,model}$, by  
 \be
F_{J,source}=\frac{R_A^2}{D^2}F_{J,model}\nn
\ee
where $R_A$ is the stellar radius and $D$ the distance to the source. Hence, the model atmosphere-dependent radius is given by 
\be
\label{Req}
R_A=D\left(\frac{F_{J,source}}{F_{J,model}}\right)^{1/2} \nn
\ee
and the model atmosphere-dependent stellar bolometric luminosity is given by
\be
L_A \equiv 4\pi R_A^2 \sigma T_A^4 = \left(4\pi D^2 F_{J,source}\right) \left(\frac{\sigma T_A^4}{F_{J,model}}\right) . \nn
\ee
Knowing \AAJ\, for an object, we deredden its observed $J$-band flux to obtain $F_{J,source}$; further knowing its mean distance $D$ (162\,pc for Cha I and 140\,pc for Taurus), and \AT\, and $F_{J,model}$ given the object's spectral type, we derive its radius $R_A$ and luminosity \AL\ using equations (3) and (4) respectively.

Note that the second quantity within parentheses, in the last equality of equation (4), corresponds to a spectral type- (i.e., \AT-) dependent synthetic bolometric correction. In Fig.\,7, and Table 3, we compare this model atmosphere bolometric correction, denoted by $BC_{J,A}$, to the empirical one used to calculate \LL\ (K. Luhman, private communication), denoted by $BC_{J,L}$. We see that, while there is a mild, non-monotonic variation between the two as a function of spectral type, they are very close in absolute value, differing by $\leq$\,0.2\,mag over the entire M spectral-type range. 

\subsection{Comparison of Model Atmosphere and Evolutionary Predictions, for Cha I and Taurus}
We now have two independent estimates for \Teff, luminosities and radii, from atmospheric and evolutionary models respectively, for all our sources. We compare these in various parameter planes. 
 
\subsubsection{Temperatures, Luminosities and Radii}
Figs\,8(a) and (b) show, for Cha I and Taurus respectively, the luminosity ratios $L_L/L_A$ vs the \Teff\ ratios $T_L/T_A$; $L_L/L_A$ vs the radius ratios $R_L/R_A$; and $T_L/T_A$ vs $R_L/R_A$.  Early, mid and late M dwarfs are depicted as blue, green and red filled circles respectively. The salient results are as follow:

\noindent {\it (1)} As discussed in \S4.2, the \AT\ derived from the model atmospheres are consistent with the evolutionary track predictions \LT\ only for mid-M types, and significantly lower for both early and late M types. $T_L/T_A$ is thus $>$1 for all early and late M sources in Cha I and Taurus in the left and right panels of Figs.\,8(a,b). 

\noindent {\it (2)} As shown in Figs.\,6(a,b) and 7, the extinctions and bolometric corrections we use to derive $L_A$ (\AAJ\ and $BC_{J,A}$) are very similar to those adopted in deriving $L_L$ (\LAJ\ and $BC_{J,L}$), with no large systematic offsets. Consequently, we expect the $L_A$ and $L_L$ values to be quite similar. This is borne out by our results in Figs.\,8(a,b), where the $L_L/L_A$ ratios are distributed roughly evenly around unity. The largest deviations are for extinction-flagged objects (where \AAJ, \LAJ\ or both are likely erroneous), with the rest agreeing to within 20\% in most cases, and deviating by at most 40\%. There also appears to be a slight spectral type variation. These trends are easily understood quantitatively from our extinction and bolometric correction plots in Figs.\,6 and 7. Our \AAJ\ vary stochastically around \LAJ\ by $\sim$0.2\,mag on average. On the other hand, there is a small spectral type-dependence in $BC_{J,L}$ versus our $BC_{J,A}$: $BC_{J,L}$ is lower by up to 0.2\,mag in the early Ms, higher by up to 0.1\,mag in the mid-Ms, and both lower by up to 0.2\,mag and higher by up to 0.1\,mag in the late Ms. Combining the random extinction errors with the small systematic undulations in $BC_J$, we expect $L_L/L_A$ to to tend to range over $\sim$1--1.4, $\sim$0.7--1.1, and $\sim$0.7--1.4 in the early, mid- and late M types respectively, which is roughly what we see.  The bottom line, thus, is that the atmospheric models predict luminosities fairly consistent (to within 30-40\%) with evolutionary expectations for PMS M types. 

\noindent {\it (3)} With no large (albeit some) systematic offset in the two luminosity estimates with M sub-type, but $T_L/T_A > 1$ in both the early and late M types, we expect the radii predicted by the evolutionary tracks, $R_L$, to be systematically lower than the $R_A$ inferred from the model atmospheres for the early and late Ms. This is clearly seen in the middle and right panels of Figs.\,8(a,b), where the majority of early and late M sources have $R_L/R_A < 1$ (even many with $L_L/L_A > 1$, which should partially compensate for $T_L/T_A > 1$).    

\subsubsection{HR Diagram}
Finally, in Figs.\,9(a,b), we plot our Cha I and Taurus sources on an HR diagram constructed from the BCAH98/CBAH00 evolutionary tracks. Black filled circles mark the location of sources using the `Luhman' parameters [$L_L$, $T_L$], while red filled circles are for our model atmosphere predictions [$L_A$, $T_A$]. 

We see that, in both Cha I and Taurus (Figs.\,9(a) and (b) respectively), using model atmosphere parameters [$L_A$, $T_A$] makes the early and late M sources appear both cooler, and hence less massive, as well as younger, than with [$L_L$, $T_L$]. This is mainly because of the difference between $T_A$ and $T_L$: the model atmosphere temperatures $T_A$ are lower than the evolutionary model predictions $T_L$ for these spectral types, and this shift also forces these sources above the theoretical isochrones, since the latter slope downwards with decreasing \Teff\ (the 30-40\% scatter between $L_L$ and $L_A$ translates to a $\sim \pm$0.15 vertical jitter on the logarithmic luminosity scale shown here, which is generally considerably less important than the vertical offset from the tracks induced by the systematic \Teff\ differences). The age discrepancy is strongest at $\gtrsim$M7, i.e., for the late M types, since the 1-few Myr isochrones descend most steeply with \Teff\ at these types. This result is somewhat less robust in Cha I, where there are relatively few late-type objects and the latest types also have large uncertainties in extinction. The effect is very clear in Taurus, however, with many more late M sources. We see that the majority of late M sources appear younger than 1 Myr when plotted with model atmosphere parameters [$L_A$, $T_A$], but fall on or below this isochrone when [$L_L$, $T_L$] are adopted instead.  Note that a handful of objects across the entire M spectral class appear above the 1 Myr isochrone even using [$L_L$, $T_L$]; presumably, these really are younger than the rest of the Taurus sample. We have also made clear on the HR diagrams which objects are known binaries/multiples, in order to rule out added emission from multiple components causing a true rise above the tracks in the late Ms. The key point, however, is that there is a {\it systematic, spectral type-dependent} offset to younger ages when the model atmosphere values [$L_A$, $T_A$] are used: the early and (especially) late M types appear younger than the mid-M sample in this case.        

These results are not unexpected. The spectral type-\Teff conversion scale $T_L$ is explicitly constructed to enforce mean coevality for members of a given star-forming region when comparing to the BCAH98/CBAH00 theoretical isochrones (\S2.2), so it is unsurprising that our Cha I and Taurus sample follow the overall shape of these isochrones (and suggest a mean age of $\sim$1--2 Myr for these regions) when $T_L$ is adopted. Moreover, a systematic drift to younger ages with later type, compared to the same isochrones, is observed when a dwarf temperature scale is used instead for M-type PMS sources, where the latter scale is cooler than $T_L$ (indeed, $T_L$ was devised to avoid this drift). Since the model atmosphere-based scale $T_A$ is also cooler than $T_L$ for the early and (particularly) the late Ms, it produces the same systematic age shift as well. The question that remains is {\it why} the $T_L$ and $T_A$ conversion scales differ; we address this below. 

\section{Discussion}
We have shown that, in general, the AMES-Dusty models outperform the newer BT-Settl ones at fitting the NIR/MIR shape. We have also found that the atmospheric models imply that both early and late M PMS sources are significantly cooler than suggested by the PMS spectral type-\Teff\ conversion scale devised by \citet{Luhman03}. We have then used the synthetic atmospheres to derive stellar properties for a large sample of diskless PMS objects, independent of the theoretical evolutionary tracks. Since the Luhman \Teff\ scale is based on these tracks, the discrepancy between the latter scale and the one we infer from the atmospheric models automatically leads to a divergence between the properties we derive and those predicted by the BCAH98/CBAH00 evolutionary models. This begs two questions: Are these discrepancies mainly a result of flaws in the atmospheric or evolutionary models? And is the cause of the disagreement in the early M types the same as in the late Ms? We review various possible answers, and their plausibilities, in \S6.1-6.3 below. In so doing, we also touch upon (in \S6.3) various reasons for the difference in performance between AMES-Dusty and BT-Settl models. 

\subsection{Uncertainties in the Template Colors}
We first consider whether the fault may lie with the L2010 template colors, instead of with either the atmospheric or evolutionary models. These templates have been derived by fitting a blue envelope to the distribution of IR colors for a large sample of young PMS sources, with the expectation that the bluest objects are those least affected by excess disk emission and/or line-of-sight dust extinction, and thus representative of naked PMS photospheres. However, the sources used are in star-forming regions, and thus even the bluest ones may have some residual contribution from disks or extinction, leading to spurious \Teff\ estimates from comparisons to synthetic atmospheres. While some residual reddening may indeed remain unaccounted for in these templates, however, one expects such contamination to be largely independent of spectral type: there is no reason for this effect to be prominent in both the early and late M templates, and absent at mid M. The fact that a discrepancy between the atmospheric and evolutionary \Teff\ is seen in the early and late M types, but not at mid M, argues strongly against spectral template errors being primarily responsible.    

\subsection{Uncertainties in the Theoretical Evolutionary Tracks}
Next, we examine whether the \Teff\ discrepancy can arise from errors in the evolutionary models. As Fig.\,5 shows, temperatures derived from the synthetic atmospheres (\AT) are lower than those based on the evolutionary models (\LT): by upto 300\,K in the early M types, and upto 500\,K at late M. Theory suggests that objects with strong magnetic fields and stellar activity -- parameters not included in these evolutionary tracks -- may indeed appear cooler than predicted by the tracks \citep[e.g.,][]{Chabrier07,MacDonald09,MacDonald12}.  This phenomenon has been invoked to explain observations of 2MASS 0535-05 (a substellar M-type PMS eclipsing binary), in which the primary is cooler than the track predictions by $\sim150$ K \citep{Stassun06,Stassun07,Mohanty09,Mohanty12}. Similarly, studies of field M dwarfs show that active objects are cooler than inactive ones of the same mass and luminosity by $\sim$50-100 K \citep{Morales08,Morales10}. However, while these \Teff\ offsets go in the right direction, they are significantly smaller than the 300--500\,K discrepancies we find in the early and late Ms. More critically, this does not explain why we find no temperature offsets at mid M, when these spectral types are just as active (if not even more) than the early and late Ms. We conclude that while magnetic field and activity effects are likely to cause some uncertainty in the temperatures predicted by the evolutionary tracks, they cannot explain either the magnitude or, especially, the spectral type-dependent trend in the \Teff\ discrepancy between the atmospheric and evolutionary models. 

\subsection{Uncertainties in the Synthetic Spectra}
Finally, we consider the scenario where the template colors are reasonably accurate, and so are the theoretical evolutionary tracks (so that \LT\, is the correct temperature at any given spectral type), with the \Teff\ discrepancies caused instead by errors in the PMS atmospheric modeling. Indeed, a priori, the fact that the trends in the synthetic \JK\ colors, and their deviations from the observed template colors, are very similar in the PMS and field Main Sequence phases (see discussion in \S4.1.2), strongly implies that the fault lies with the synthetic atmospheres in both cases (since the evolutionary stages in the two cases are very different, and reddening uncertainties are negligible for the field templates).  

We start by considering the early M types (M0--M2). It is useful in this regard to additionally consider the \JK\ colors for the mid to late K dwarfs, also plotted in Fig.\,1. As discussed in \S4.1.2, the $\sim$0.075\,mag deviation between the model (at $T_L$) and template \JK\ at M0 persists into the K types, while below M0, and down to M2, the deviation decreases. Our detailed modeling in \S4.2.1 moreover reveals that while AMES-Dusty models provide the best fit overall to the early M spectral types, the implied \Teff\ is much lower than \LT; at \LT\, itself, BT-Settl yields a much better fit, with small deviations only at $J$ and 8.0\,$\mu$m. 

We propose that these trends can be explained by the treatment of convection. Specifically, the theoretical evolutionary models imply that spectral types $\sim$M3 and earlier (\LT\, $\gtrsim$ 3400\,K) correspond to masses $\gtrsim$ 0.4\,\Msun\ at ages of a few Myr (see tracks in Figs.\,9(a,b)). At such young ages (and hence low gravities), these masses evince extended super-adiabatic layers during PMS evolution, and their modeling is thus very sensitive to the adopted treatment of convection (i.e., to the mixing length $l_{mix}$, within the Mixing Length Theory (MLT) formalism adopted in all the models considered here; \citealt{Baraffe02}). We thus consider it plausible that (small) adjustments to $l_{mix}$ can reconcile the atmospheric models and data for spectral types $\lesssim$ M3; indeed, for AMES-Cond/Dusty $l_{mix}=H_p$, whereas for BT-Settl $l_{mix}=2H_p$, which may help explain the better fits at \LT\ for BT-Settl. Below M3, the super-adiabatic layers begin to retract and the modeling becomes relatively insensitive to $l_{mix}$ by M4, bringing the synthetic atmospheres into better agreement with the spectral templates by the mid M types at the `correct' temperature $T_L$.

Next, in the mid M types (M3--M6), our detailed analysis in \S4.2.1 shows that AMES-Dusty yields good fits to the templates at \Teff\ very close to the evolutionary model-predicted temperatures \LT. It is suggestive that M3 corresponds to \LT\, $\sim$ 3400\,K, just about where H$_2$O becomes the principal source of opacity in the IR \citep{Allard94,Allard95} and convective effects become less important. Interestingly, in spite of the improved elemental abundances in BT-Settl, the latter models yield worse fits than AMES-Dusty at \LT\, ({\it and} poorer quality best-fits than AMES-Dusty even without constraining the temperature to \LT) over M3--M6 (see \S4.2.2 and Fig.\,3). This is possibly due to differences in the H$_2$O linelists and opacities between the two models: Dusty incorporates the AMES list from \citet{Partridge97}, while Settl is based on the BT2 list from \citet{Barber06}. Empirically, compared to AMES-Dusty and the data, BT-Settl under-predicts the flux at $J$ and over-predicts the flux longwards of 3.6\,$\mu$m at \LT, for M3--M6 (Fig.\,3). At the same time, tests indicate that the BT2 H$_2$O opacities are significantly more complete than AMES in the $J$ band \citep{Barber06,Lyubchik07}; both sets, however, still appear incomplete at longer wavelengths (specifically, at $K$; \citealt{Allard12b}). We therefore speculate that the improved (higher) H$_2$O opacity in $J$ with BT2, combined with remaining missing opacities for this molecule at longer wavelengths, makes the BT-Settl models appear slightly too red compared to the M3--M5 templates at \LT\, by allowing too much flux to escape in the MIR, while the lacks in the AMES H$_2$O opacities at {\it both} $J$ and longer wavelengths conspire to give spuriously better fits to AMES-Dusty at \LT\, for these spectral types (a similar effect has been noted with the incomplete H$_2$O linelists employed in the NextGen models; \citealt{Allard12b}). This needs to be verified in future studies.

Finally, in the late M types (M7--L0), we find that both AMES-Dusty and BT-Settl models are too blue compared to the spectral templates at \LT, with the deviation increasing with later type. The only new and significant source of opacity that appears in this spectral type range is dust, which acts to redden the spectrum: both via a suppression of flux at shorter wavelengths due to grain opacity, and an enhancement of flux at longer wavelengths due to a reduction in H$_2$O opacity, caused by H$_2$O destruction through grain-induced backwarming. It is therefore no great leap to suggest that the atmospheric models currently underestimate the dust opacity at these spectral types -- either by having too little dust formation at a given \Teff, and/or by inadequate modeling of grain size and shape (e.g., \citealt{Allard12b} suggest large porous grains may be required to explain a similar discrepancy in late-type field M dwarfs).  

The only apparent impediment to this solution is that grains are usually stated to become {\it important} in these synthetic atmospheres at \Teff\,$\lesssim$ 2600\,K (e.g., \citealt{Allard12b}), while we require them to become evident at a higher temperature of $\sim$2800\,K, in order to explain the deviation between atmospheric models and spectral templates by $\sim$ M7 (assuming that \LT\ is the physically correct PMS temperature at these types). However, a close look at Fig.\,1 shows that the AMES-Cond and AMES-Dusty PMS models begin to diverge by $\sim$3000\,K, with Dusty rapidly becoming redder at lower \Teff; the Cond and Dusty Main Sequence models (at much higher gravity) also similarly diverge at the same temperature \citep{Allard12b}. Since the only salient difference between AMES-Cond and AMES-Dusty is in their treatment of dust opacity (neglected in the former but included in the latter), this suggests that {\it some} dust starts to form in these synthetic atmospheres by 3000\,K, though it is not yet a strong opacity source. We simply propose that the opacity of these grains be enhanced, either through an increased dust formation efficiency, and/or changes to the grain structure and geometry.

\section{Conclusions}
To summarise, our primary conclusions are as follow:

\noindent $\bullet$ From fits to NIR photometry over 1.2--24\,$\mu$m, state-of-the-art model atmospheres imply effective temperatures $T_A$ for early and late PMS M types that are significantly lower than the evolutionary model predictions $T_L$ for these types: by up to 300\,K for the early Ms (M0--M2), and up to 500\,K for the late Ms (M7--L0). The two \Teff\ estimates agree only in the mid M types.  

\noindent $\bullet$ Conversely, the luminosities implied by the synthetic atmospheres for M type PMS sources are consistent with empirically determined values, to within 30--40\%.

\noindent $\bullet$ The above trends in \Teff\ and luminosities cause early and (especially) late M type PMS objects to appear systematically cooler, less massive, larger and younger when placed on theoretical HR diagrams using the model atmosphere-derived parameters, compared to evolutionary predictions.

\noindent $\bullet$ While magnetic field/activity effects can cause \Teff\ in M types to be lower than evolutionary models predict, this cannot explain the magnitude or the spectral type-dependency of the above \Teff\ discrepancy. 

\noindent $\bullet$ We propose that the discrepancy arises due to errors in the synthetic spectra. Specifically, among the early M types ($M_{\ast} \gtrsim 0.4$\,\Msun), it is due to deficiencies in the atmospheric modeling of convection, while in the late M types, it is due to an underestimation of dust opacity (and an attendant overestimation of H$_2$O opacity due to insufficient backwarming). The good agreement in the mid Ms suggests that the outdated H$_2$O opacities in the older model atmospheres fortuitously match observations when convection becomes less important and grain opacity effects are negligible.

\noindent $\bullet$ The above explanation requires that dust opacity contribute significantly by $\sim$M7 (2800\,K), i.e., at a \Teff\ $\sim$200\,K higher than predicted by the current crop of synthetic atmospheres. This may be accomplished by increasing the grain formation efficiency, and/or better modeling of grain size, shape and structure. 

\noindent $\bullet$ BT-Settl atmospheric models overall give worse best-fits to the data than the earlier AMES-Dusty models, when the model temperature is allowed to vary. However, this is not necessarily an indictment of the newer (and thus presumably better) elemental abundances and H$_2$O opacities in BT-Settl. In particular, BT-Settl gives better fits than AMES-Dusty to early M types, when the \Teff\ is fixed at the $T_L$ predicted by the evolutionary tracks for these types (which we argue is the `correct' temperature); we ascribe this to possibly better treatment of convection, in the form of the mixing length scale. Similarly, we argue that the worse fits to BT-Settl at $T_L$ for mid M may in fact result from the improvements in its $J$-band H$_2$O opacity, coupled with remaining lacks in this opacity at longer wavelengths (while the better fits here to AMES-Dusty may be spurious, arising from opacity errors at both $J$ and longer wavelengths roughly cancelling out). At late M, however, the lacks in grain opacity discussed above are present in both BT-Settl and AMES-Dusty.  

Finally, we note that various analyses demand a good model for the shape of the stellar photospheric SED (e.g., for extracting the IR disk excess for a young star). The best fit atmospheric models we have found in this work, while not necessarily at the correct {\it temperature} for a given M spectral type, are very good matches to the NIR SED shapes of M types, and can be profitably employed for such modeling purposes.  

\newpage

\bibliographystyle{apjliketottle}
\bibliography{bibliography}

\newpage

\setcounter{figure}{5}
\makeatletter 
\renewcommand{\thefigure}{\@arabic\c@figure a}
\makeatother
\begin{figure}[h!]
\begin{center}
\includegraphics[width=10cm]{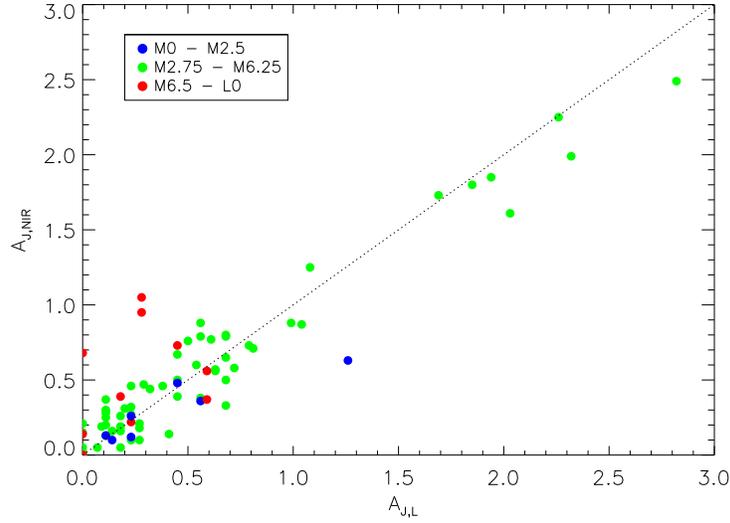}
\caption{The extinctions measured for the 78 objects in our Cha I sample (Table 1) derived using the method described in \S5.1.1, \AAJ, plotted against their optically-dervied extinctions \citep{Luhman07}, \LAJ. We group the data into spectral type bins, as provided in the legend. No systematic offset, nor any dependency on spectral type, appears to exist.}
\label{LAJvAAJ}
\end{center}
\end{figure}

\setcounter{figure}{5}
\makeatletter 
\renewcommand{\thefigure}{\@arabic\c@figure b}
\makeatother
\begin{figure}[h!]
\begin{center}
\includegraphics[width=10cm]{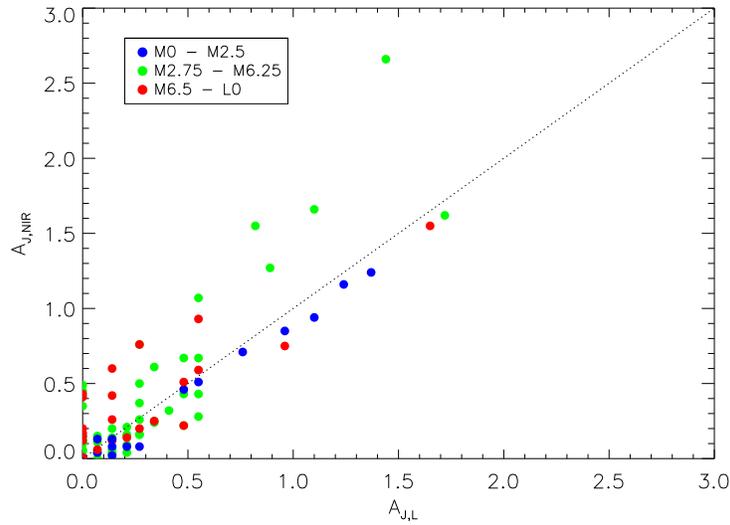}
\caption{Same as Fig.\,\ref{LAJvAAJ}, but for the Taurus sample (Table 2).}
\label{LAJvAAJTaurus}
\end{center}
\end{figure}
\clearpage

\makeatletter 
\renewcommand{\thefigure}{\@arabic\c@figure}
\makeatother
\begin{figure}[h!]
\begin{center}
\includegraphics[width=10cm]{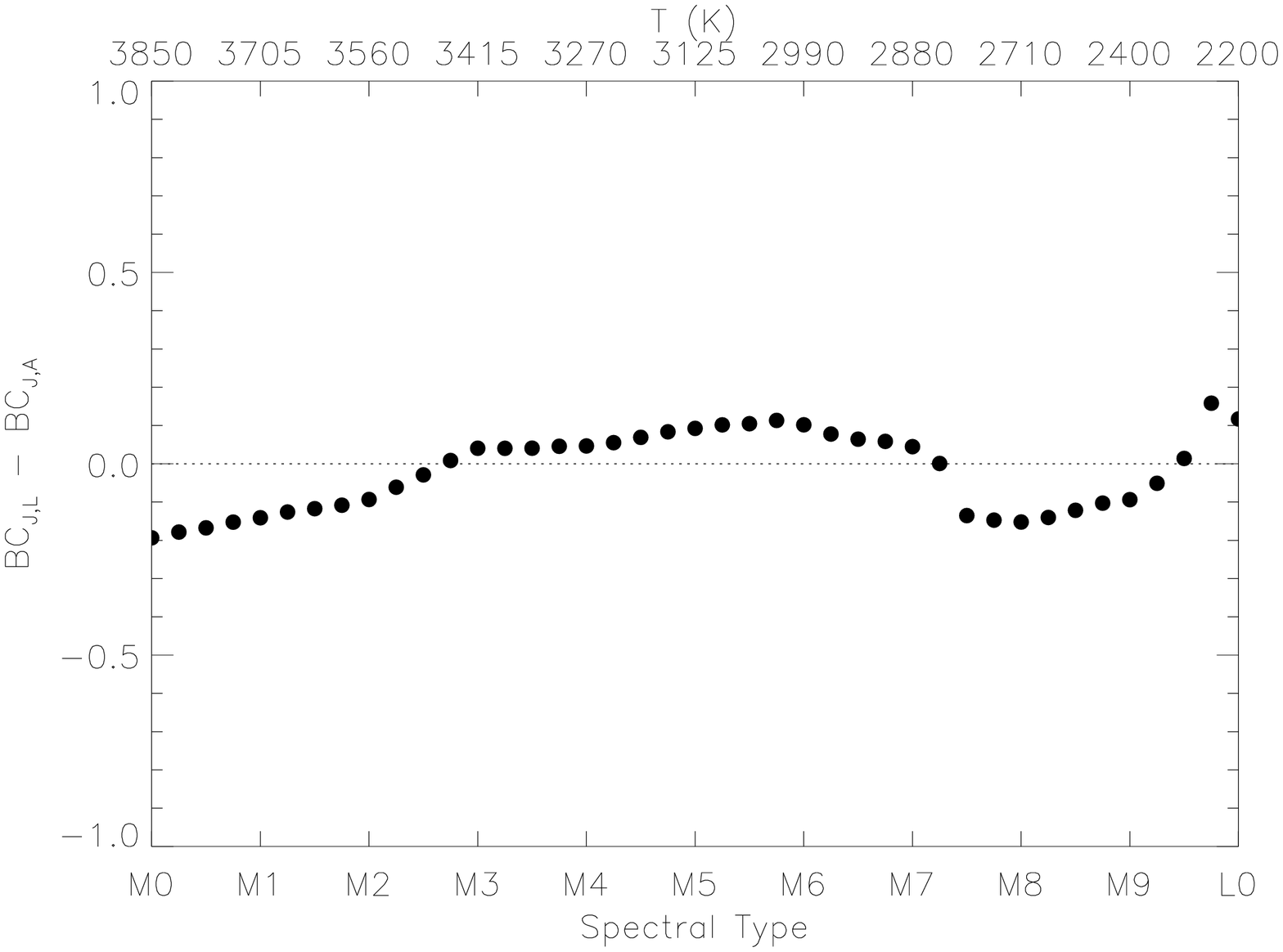}
\caption{Bolometric correction differences between those used by Luhman (empirical estimates) and those calculated from the best-fitting synthetic spectra in this paper, showing the similarity between the two across the M types. Explicit values for each spectral subclass are given in Table 3.}
\label{BCplot}
\end{center}
\end{figure}

\makeatletter 
\renewcommand{\thefigure}{\@arabic\c@figure a}
\makeatother
\begin{figure}[h!]
\begin{center}
\includegraphics[width=10cm]{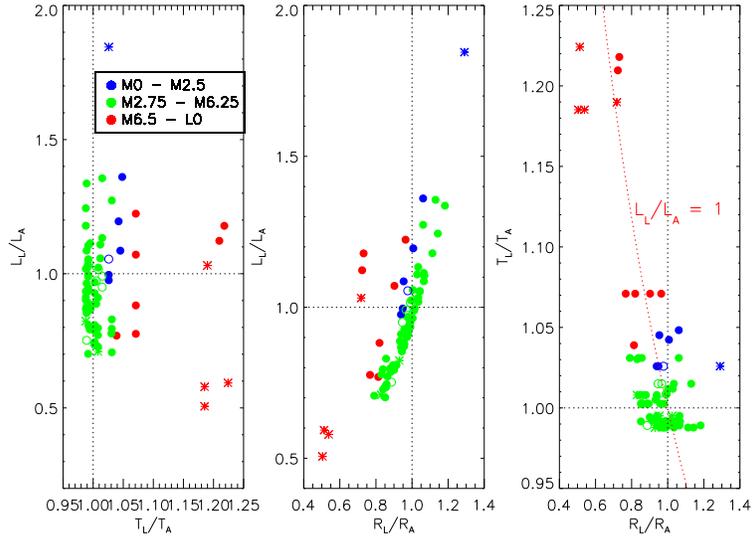}
\caption{Ratios of the Luhman and Allard radii, luminosities, and temperatures for our Cha I sample, plotted against eachother. Asterisked data points mark those objects which have been flagged, either due to having a poor \AJ \ fit or a larger-than-expected deviation from the originals supplied in \citet{Luhman07}, whilst open circles represent known binaries \citep{Lafreniere08}. Objects have been grouped into early, mid, and late M type bins, and a line indicating equal luminosity has been included in the rightmost plot to reinforce the point in Fig.\,\ref{BCplot}: a large value of $T_L/T_A$ is usually balanced by a small $R_L/R_A$ to produce similar values of $L_L$ and $L_A$, due to the comparable bolometric corrections.}
\label{RTLplots}
\end{center}
\end{figure}

\setcounter{figure}{7}
\makeatletter 
\renewcommand{\thefigure}{\@arabic\c@figure b}
\makeatother
\begin{figure}[h!]
\begin{center}
\includegraphics[width=10cm]{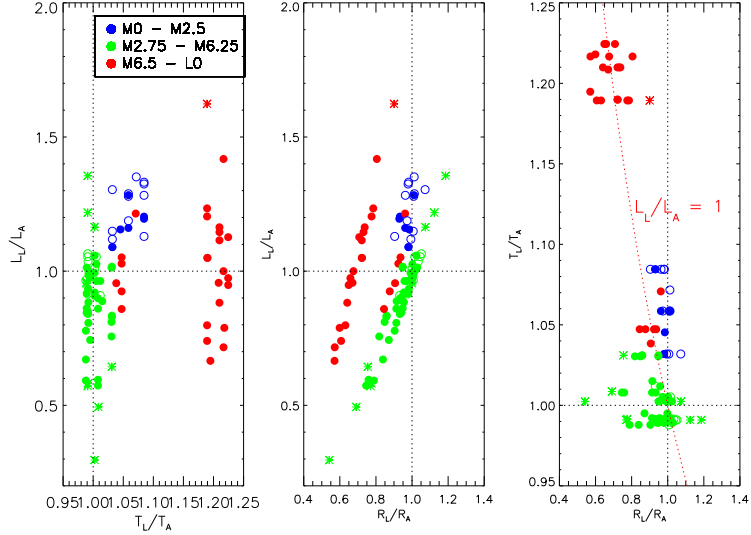}
\caption{Same as Fig.\,\ref{RTLplots}, but for the Taurus sample. Binary information taken from \citet{Kraus11}.}
\label{RTLplotsT}
\end{center}
\end{figure}

\makeatletter 
\renewcommand{\thefigure}{\@arabic\c@figure a}
\makeatother
\begin{figure}[h!]
\begin{center}
\includegraphics[width=10cm]{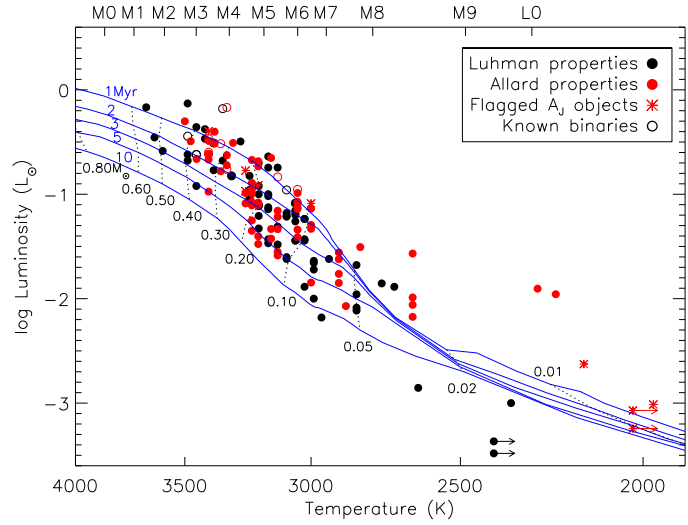}
\caption{H-R diagram for the 78 Cha I objects in our sample, plotted for both Luhman (black) and Allard (red) properties. Asterisked data points mark the flagged objects due to the \AJ \ fitting, whilst open circles represent known binaries/multiples. The \citet{Luhman03} spectral type-\Teff \ relation is included on the top axis, and the BCAH98/CBAH00 theoretical evolutionary tracksare indicated in blue, with ages (in Myr) and masses (in \Msun) specified.}
\label{HRdiagram}
\end{center}
\end{figure}

\setcounter{figure}{8}
\makeatletter 
\renewcommand{\thefigure}{\@arabic\c@figure b}
\makeatother
\begin{figure}[h!]
\begin{center}
\includegraphics[width=10cm]{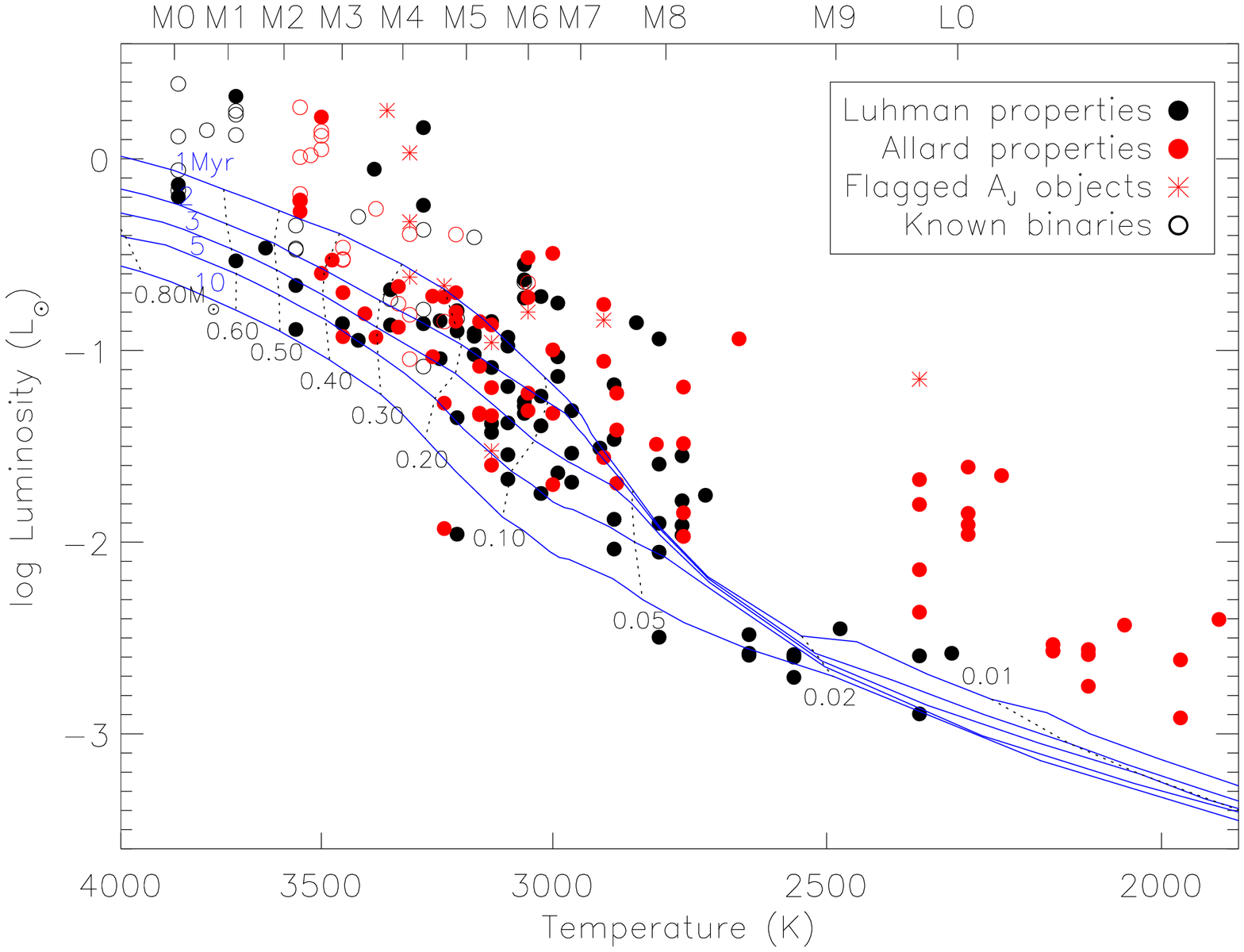}
\caption{Same as Fig.\,\ref{HRdiagram}, but for the Taurus sample.}
\label{HRdiagramT}
\end{center}
\end{figure}

\clearpage
\newpage

\scriptsize 
\begin{longtable}{lllllllllll}

\caption[Properties of the 78 Class III Cha I objects]{Properties of the 78 Class III Cha I objects} \label{78objects} \\

\hline \hline \multicolumn{1}{c}{Name} & \multicolumn{1}{c}{Spec Type} & \multicolumn{1}{c}{$J$} & \multicolumn{1}{c}{$H$} & \multicolumn{1}{c}{$K$} & \multicolumn{1}{c}{\LAJ} & \multicolumn{1}{c}{\LT(K)} & \multicolumn{1}{c}{\LL(\Lsun)} &  \multicolumn{1}{c}{\AAJ} & \multicolumn{1}{c}{\AT(K)} & \multicolumn{1}{c}{\AL(\Lsun)} \rule{0pt}{2.6ex} \rule[-1.2ex]{0pt}{0pt} \\ \hline 
\endfirsthead

\multicolumn{11}{c}%
{{ \tablename\ \thetable{} -- \textit{continued}}} \\
\hline \hline \multicolumn{1}{c}{Name} & \multicolumn{1}{c}{Spec Type} & \multicolumn{1}{c}{$J$} & \multicolumn{1}{c}{$H$} & \multicolumn{1}{c}{$K$} & \multicolumn{1}{c}{\LAJ} & \multicolumn{1}{c}{\LT(K)} & \multicolumn{1}{c}{\LL(\Lsun)} &  \multicolumn{1}{c}{\AAJ} & \multicolumn{1}{c}{\AT(K)} & \multicolumn{1}{c}{\AL(\Lsun)} \rule{0pt}{2.6ex} \rule[-1.2ex]{0pt}{0pt} \\ \hline 
\endhead

J10523694-7440287 & M4.75 & 11.45 & 10.71 & 10.44 & 0.18 & 3161 & 0.10 &	0.26$^{\mathrm{a}}$ & 3200 & 0.12 \\
J11011370-7722387 & M5.25 & 13.06 & 12.38 & 12.09 & 0.27 & 3091 & 0.025 &	0.21 & 3125 & 0.026 \\
J11011926-7732383 & M7.25 & 13.41$^{\mathrm{c}}$ & 12.57$^{\mathrm{c}}$ & 11.97$^{\mathrm{c}}$ & 0.45 & 2838 & 0.021 &	0.73 & 2650 & 0.027 \\
J11021927-7536576 & M4.5 & 12.13 & 11.54 & 11.19 & 0.27 & 3198 & 0.062 &	0.10 & 3225 & 0.056 \\
J11022610-7502407 & M4.75 & 11.76 & 11.11 & 10.81 & 0.14 & 3161 & 0.075 &	0.16 & 3200 & 0.083 \\
J11024183-7724245 & M5 & 12.80 & 11.99 & 11.61 & 0.72 & 3125 & 0.049 &	0.58 & 3150 & 0.047 \\
J11034186-7726520 & M5.5 & 13.00 & 12.11 & 11.69 & 0.61 & 3058 & 0.036 &	0.77 & 3050 & 0.046 \\
J11034764-7719563 & M5 & 11.31 & 10.41 & 9.99 & 0.68 & 3125 & 0.18 &	0.79 & 3150 & 0.22 \\
J11035682-7721329 & M3.5 & 10.80 & 9.99 & 9.71 & 0.20 & 3342 & 0.21 &	0.31 & 3325 & 0.24 \\
J11041060-7612490 & M6 & 13.16 & 12.52 & 12.12 & 0.11 & 2990 & 0.019 &	0.25 & 2900 & 0.024 \\
J11045285-7625514 & M1.75 & 10.72 & 9.98 & 9.75 & 0.23 & 3596 & 0.26 &	0.12 & 3450 & 0.22 \\
J11051467-7711290$^{\mathrm{e}}$ & M3.25 & 10.94 & 10.03 & 9.61 & 0.79 & 3379 & 0.31 &	0.73 & 3350 & 0.31 \\
J11052472-7626209 & M2.75 & 11.47 & 10.74 & 10.52 & 0.23 & 3451 & 0.12 &	0.10 & 3400 & 0.11 \\
J11054300-7726517$^{\mathrm{e}}$ & M5.25 & 11.26 & 10.62 & 10.23 & 0.11 & 3091 & 0.11 &	0.28 & 3125 & 0.15 \\
J11055261-7618255 & M1.5 & 10.31 & 9.59 & 9.34 & 0.11 & 3632 & 0.35 &	0.13 & 3475 & 0.32 \\
J11060010-7507252 & M4.5 & 12.33 & 11.75 & 11.42 & 0.18 & 3198 & 0.047 &	0.05 & 3225 & 0.045 \\
J11061545-7737501 & M2.75 & 12.65 & 10.99 & 10.25 & 2.82 & 3451 & 0.44 &	2.49 & 3400 & 0.32 \\
J11062877-7737331 & M3.25 & 12.67 & 11.32 & 10.70 & 1.85 & 3379 & 0.17 &	1.80 & 3350 & 0.17 \\
J11063799-7743090 & M6.5 & 12.97 & 12.26 & 11.81 & 0.18 & 2935 & 0.024 &	0.39 & 2825 & 0.031 \\
J11064346-7726343 & M3 & 10.81 & 9.79 & 9.39 & 0.99 & 3415 & 0.42 &	0.88 & 3375 & 0.40 \\
J11065733-7742106 & M4.25 & 11.44 & 10.51 & 10.21 & 0.54 & 3234 & 0.15 &	0.60$^{\mathrm{a}}$ & 3250 & 0.17 \\
J11070324-7610565 & M6 & 13.85 & 13.12 & 12.75 & 0.11 & 2990 & 0.010 &	0.37 & 2900 & 0.014 \\
J11071148-7746394 & M3 & 11.08 & 10.08 & 9.66 & 1.04 & 3415 & 0.34 &	0.87 & 3375 & 0.31 \\
J11071915-7603048 & M2.5 & 10.96 & 10.09 & 9.77 & 0.45 & 3488 & 0.24 &	0.48 & 3400 & 0.24 \\
J11072040-7729403 & M4.5 & 11.13 & 10.55 & 10.26 & 0.00 & 3198 & 0.12 &	0.00 & 3225 & 0.13 \\
J11072443-7743489 & M5.75 & 13.45 & 12.42 & 11.84 & 1.08 & 3024 & 0.036 &	1.25 & 3000 & 0.047 \\
J11072647-7742408 & $\geq$M9 & 17.53$^{\mathrm{d}}$ & 16.45$^{\mathrm{d}}$ & 15.59$^{\mathrm{d}}$ & 0.28 & $\leq$2400 & 0.00043 &	1.11$^{\mathrm{a,b}}$ & $\leq$2025 & 0.00085 \\
J11073519-7734493 & M4.25 & 12.13 & 11.28 & 10.95 & 0.68 & 3234 & 0.091 &	0.50 & 3250 & 0.082 \\
J11073686-7733335 &$^{\mathrm{e}}$ M3.5 & 11.59 & 10.05 & 9.35 & 2.26 & 3342 & 0.66 &	2.25 & 3325 & 0.68 \\
J11073775-7735308 & M7.75 & 13.61 & 12.90 & 12.42 & 0.23 & 2752 & 0.014 &	0.22 & 2275 & 0.012 \\
J11073832-7747168 & M4.5 & 12.24 & 11.40 & 11.03 & 0.63 & 3198 & 0.078 &	0.57 & 3225 & 0.078 \\
J11073840-7552519 & M4.75 & 12.77 & 12.12 & 11.80 & 0.27 & 3161 & 0.034 &	0.18 & 3200 & 0.033 \\
J11074610-7740089 & M5.75 & 12.78 & 12.00 & 11.51 & 0.45 & 3024 & 0.037 &	0.67 & 3000 & 0.051 \\
J11075225-7736569 & M5.5 & 12.29 & 11.52 & 11.10 & 0.63 & 3058 & 0.070 &	0.56 & 3050 & 0.072 \\
J11075993-7715317 & M5.75 & 12.52 & 11.65 & 11.17 & 0.68 & 3024 & 0.058 &	0.80 & 3000 & 0.073 \\
J11080234-7640343 & M6 & 12.94 & 12.31 & 11.94 & 0.11 & 2990 & 0.023 &	0.20 & 2900 & 0.028 \\
J11081648-7744371 & M3.75 & 11.20 & 10.34 & 10.02 & 0.29 & 3306 & 0.15 &	0.47 & 3325 & 0.19 \\
J11081703-7744118 & M5.5 & 11.79 & 11.06 & 10.67 & 0.32 & 3058 & 0.083 &	0.44 & 3050 & 0.10 \\
J11081896-7739170 & M5.5 & 12.15 & 11.42 & 11.02 & 0.23 & 3058 & 0.055 &	0.46 & 3050 & 0.075 \\
J11082404-7739299 & M6.25 & 14.31 & 13.58 & 13.24 & 0.11 & 2962 & 0.0066 &	0.30 & 2875 & 0.0085 \\
J11082410-7741473$^{\mathrm{e}}$ & M5.5 & 12.05 & 11.20 & 10.71 & 0.56 & 3058 & 0.082 &	0.79 & 3050 & 0.11 \\
J11083040-7731387 & $\geq$M9 & 17.84$^{\mathrm{d}}$ & 16.75$^{\mathrm{d}}$ & 15.97$^{\mathrm{d}}$ & 0.28 & $\leq$2400 & 0.00033 &	0.99$^{\mathrm{b}}$ & $\leq$ 2025 & 0.00057 \\
J11084069-7636078 & M2.5 & 10.56 & 9.66 & 9.28 & 1.26 & 3488 & 0.74 &	0.63$^{\mathrm{b}}$ & 3400 & 0.40 \\
J11085176-7632502 & M7.25 & 14.29 & 13.53 & 12.96 & 0.59 & 2838 & 0.011 &	0.56 & 2650 & 0.010 \\
J11085421-7732115 & M5.25 & 12.31 & 11.56 & 11.22 & 0.56 & 3091 & 0.066 &	0.38 & 3125 & 0.061 \\
J11085596-7727132 & M5.25 & 13.51 & 12.29 & 11.62 & 1.69 & 3091 & 0.061 &	1.73 & 3125 & 0.070 \\
J11091380-7628396 & M4.75 & 11.85 & 11.21 & 10.87 & 0.18 & 3161 & 0.072 &	0.19 & 3200 & 0.079 \\
J11092913-7659180 & M5.25 & 13.27 & 12.51 & 12.11 & 0.45 & 3091 & 0.024 &	0.50 & 3125 & 0.028 \\
J11093543-7731390 & M8.25 & 15.92$^{\mathrm{d}}$ & 14.99$^{\mathrm{d}}$ & 14.37$^{\mathrm{d}}$ & 0.00 & 2632 & 0.0014 &	0.68$^{\mathrm{b}}$ & 2150 & 0.0023 \\
J11094006-7628391$^{\mathrm{e}}$ & M1.25 & 10.07 & 9.23 & 8.96 & 0.56 & 3669 & 0.68 &	0.36 & 3500 & 0.50 \\
J11094525-7740332 & M5.75 & 12.35 & 11.45 & 11.03 & 0.50 & 3024 & 0.058 &	0.76$^{\mathrm{a}}$ & 3000 & 0.082 \\
J11094918-7731197 & M5.5 & 13.06 & 12.23 & 11.80 & 0.68 & 3058 & 0.036 &	0.65 & 3050 & 0.039 \\
J11100192-7725451 & M5.25 & 13.83 & 12.60 & 12.02 & 2.03 & 3091 & 0.062 &	1.61 & 3125 & 0.046 \\
J11100658-7642486 & M9.25 & 16.34 & 15.86 & 15.07 & 0.00 & 2350 & 0.001 &	0.00$^{\mathrm{a}}$ & 1975 & 0.00097 \\
J11101153-7733521 & M4.5 & 12.18 & 11.19 & 10.78 & 0.56 & 3198 & 0.077 &	0.88 & 3225 & 0.110 \\
J11102226-7625138 & M8 & 13.53 & 12.90 & 12.45 & 0.00 & 2710 & 0.013 &	0.00 & 2225 & 0.011 \\
J11102852-7716596 & M5.5 & 11.73 & 11.11 & 10.78 & 0.00 & 3058 & 0.066 &	0.15 & 3050 & 0.083 \\
J11103481-7722053 & M4 & 12.04 & 10.72 & 10.03 & 1.94 & 3270 & 0.32 &	1.85 & 3300 & 0.31 \\
J11103644-7722131 & M4.75 & 12.72 & 11.37 & 10.67 & 2.32 & 3161 & 0.23 &	1.99 & 3200 & 0.18 \\
J11104006-7630547 & M7.25 & 14.57 & 13.85 & 13.34 & 0.59 & 2838 & 0.0082 &	0.37 & 2650 & 0.0067 \\
J11105076-7718031 & M4.25 & 12.04 & 11.10 & 10.75 & 0.81 & 3234 & 0.11 &	0.71$^{\mathrm{a}}$ & 3250 & 0.107 \\
J11112260-7705538 & M4.5 & 11.78 & 11.00 & 10.69 & 0.45 & 3198 & 0.10 &	0.39 & 3225 & 0.10 \\
J11113474-7636211 & M2.5 & 10.86 & 10.08 & 9.80 & 0.23 & 3488 & 0.21 &	0.26 & 3400 & 0.22 \\
J11115400-7619311$^{\mathrm{e}}$ & M2.5 & 10.20 & 9.53 & 9.23 & 0.14 & 3488 & 0.36 &	0.10 & 3400 & 0.34 \\
J11120288-7722483 & M6 & 13.59 & 12.94 & 12.51 & 0.68 & 2990 & 0.022 &	0.33 & 2900 & 0.017 \\
J11120327-7637034 & M5.5 & 11.77 & 11.11 & 10.78 & 0.00 & 3058 & 0.063 &	0.21 & 3050 & 0.084 \\
J11132737-7634165$^{\mathrm{e}}$ & M2.75 & 10.61 & 9.86 & 9.63 & 0.11 & 3451 & 0.24 &	0.13 & 3400 & 0.24 \\
J11132970-7629012 & M4.25 & 11.57 & 10.86 & 10.58 & 0.09 & 3234 & 0.089 &	0.19 & 3250 & 0.10 \\
J11133356-7635374 & M4.5 & 11.64 & 11.04 & 10.73 & 0.07 & 3198 & 0.080 &	0.05 & 3225 & 0.084 \\
J11141565-7627364$^{\mathrm{e}}$ & M3.75 & 11.29 & 10.47 & 10.12 & 0.38 & 3306 & 0.15 &	0.46 & 3325 & 0.173 \\
J11142906-7625399 & M4.75 & 12.57 & 11.90 & 11.61 & 0.18 & 3161 & 0.037 &	0.16 & 3200 & 0.040 \\
J11145031-7733390$^{\mathrm{e}}$ & M2.75 & 10.48 & 9.75 & 9.55 & 0.00 & 3451 & 0.24 &	0.05 & 3400 & 0.25 \\
J11152180-7724042 & M4.75 & 11.76 & 11.13 & 10.82 & 0.41 & 3161 & 0.096 &	0.14 & 3200 & 0.081 \\
J11173792-7646193 & M5.75 & 13.51 & 12.95 & 12.62 & 0.00 & 3024 & 0.013 &	0.02 & 3000 & 0.014 \\
J11194214-7623326 & M5 & 12.73 & 12.03 & 11.72 & 0.23 & 3125 & 0.033 &	0.27 & 3150 & 0.038 \\
J11195652-7504529 & M7.25 & 14.05 & 13.33 & 12.98 & 0.00 & 2838 & 0.0077 &	0.14 & 2650 & 0.0088 \\
J11242980-7554237 & M4.75 & 10.93 & 10.20 & 9.88 & 0.23 & 3161 & 0.18 &	0.32 & 3200 & 0.21 \\
J11332327-7622092 & M4.5 & 10.59 & 10.00 & 9.71 & 0.00 & 3198 & 0.20 &	0.01 & 3225 & 0.21 \\
        \hline
\end{longtable}
\begin{minipage}{.9\linewidth}
\renewcommand{\footnoterule}{}
\footnotetext[1]{ Flagged due to the best fit \AJ \ having an rms above 1.00.}
\footnotetext[2]{ Flagged due to the best fit \AJ \ differing to the \citet{Luhman07} derived value by above 0.5.}
\footnotetext[3]{ Photometry taken from \citet{Luhman04b} due to it being an unresolved binary in 2MASS.}
\footnotetext[4]{ Photometry taken from \citet{Luhman07} from his ISPI measurements.}
\footnotetext[5]{ Known binary \citep{Lafreniere08}.}
\end{minipage}
\normalsize

\newpage

\scriptsize
\begin{longtable}{lllllllllll}

\caption[Properties of the 96 Class III Taurus objects]{Properties of the 96 Class III Taurus objects} \label{96objects} \\

\hline \hline \multicolumn{1}{c}{Name} & \multicolumn{1}{c}{Spec Type} & \multicolumn{1}{c}{$J$} & \multicolumn{1}{c}{$H$} & \multicolumn{1}{c}{$K$} & \multicolumn{1}{c}{\LAJ} & \multicolumn{1}{c}{\LT(K)} & \multicolumn{1}{c}{\LL(\Lsun)} &  \multicolumn{1}{c}{\AAJ} & \multicolumn{1}{c}{\AT(K)} & \multicolumn{1}{c}{\AL(\Lsun)} \rule{0pt}{2.6ex} \rule[-1.2ex]{0pt}{0pt} \\ \hline 
\endfirsthead

\multicolumn{11}{c}%
{{ \tablename\ \thetable{} -- \textit{continued}}} \\
\hline \hline \multicolumn{1}{c}{Name} & \multicolumn{1}{c}{Spec Type} & \multicolumn{1}{c}{$J$} & \multicolumn{1}{c}{$H$} & \multicolumn{1}{c}{$K$} & \multicolumn{1}{c}{\LAJ} & \multicolumn{1}{c}{\LT(K)} & \multicolumn{1}{c}{\LL(\Lsun)} &  \multicolumn{1}{c}{\AAJ} & \multicolumn{1}{c}{\AT(K)} & \multicolumn{1}{c}{\AL(\Lsun)} \rule{0pt}{2.6ex} \rule[-1.2ex]{0pt}{0pt} \\ \hline 
\endhead

J04034930+2610520 & M3.5 & 10.27 & 9.70 & 9.46 & 0.00 & 3342 & 0.21 & 0.00 & 3325 & 0.22 \\
J04035084+2610531 & M2 & 10.37 & 9.75 & 9.53 & 0.00 & 3560 & 0.22 & 0.00 & 3450 & 0.20 \\
J04043936+2158186 & M3.5 & 10.80 & 10.17 & 9.97 & 0.07 & 3342 & 0.14 & 0.00 & 3325 & 0.13 \\
J04043984+2158215 & M3 & 10.94 & 10.35 & 10.10 & 0.00 & 3415 & 0.11 & 0.00 & 3375 & 0.12 \\
J04053087+2151106 & M2 & 10.95 & 10.29 & 10.06 & 0.00 & 3560 & 0.13 & 0.00 & 3450 & 0.12 \\
J04131414+2819108 & M4 & 9.64 & 8.87 & 8.62 & 0.48 & 3270 & 0.57 & 0.22$^{\mathrm{a}}$ & 3300 & 0.47 \\
J04132722+2816247$^{\mathrm{d}}$ & M0 & 8.83 & 7.79 & 7.46 & 0.96 & 3850 & 2.5 & 0.85 & 3550 & 1.9 \\
J04144739+2803055 & M5.25 & 10.80 & 10.17 & 9.92 & 0.00 & 3091 & 0.12 & 0.06 & 3125 & 0.14 \\
J04144797+2752346$^{\mathrm{d}}$ & M1 & 8.36 & 7.62 & 7.42 & 0.21 & 3705 & 1.8 & 0.08 & 3500 & 1.4 \\
J04145234+2805598 & M3.25 & 9.53 & 8.21 & 7.71 & 0.82 & 3378 & 0.88 & 1.55$^{\mathrm{c}}$ & 3350 & 1.8 \\
J04150515+2808462 & M5.5 & 10.11 & 9.42 & 9.09 & 0.27 & 3057 & 0.28 & 0.26 & 3050 & 0.30 \\
J04151471+2800096 & M8.5 & 15.10 & 14.25 & 13.77 & 0.14 & 2555 & 0.0026 & 0.26 & 2100 & 0.0026 \\
J04152409+2910434 & M7 & 13.68 & 12.88 & 12.36 & 0.55 & 2880 & 0.013 & 0.59 & 2750 & 0.014 \\
J04161885+2752155 & M6.25 & 12.55 & 11.78 & 11.35 & 0.27 & 2962 & 0.029 & 0.50 & 2875 & 0.038 \\
J04162725+2053091 & M5 & 12.05 & 11.47 & 11.11 & 0.00 & 3125 & 0.037 & 0.14 & 3150 & 0.046 \\
J04163048+3037053 & M4.5 & 13.62 & 12.97 & 12.62 & 0.21 & 3197 & 0.011 & 0.21 & 3225 & 0.012 \\
J04173893+2833005$^{\mathrm{d}}$ & M2 & 9.98 & 9.29 & 9.05 & 0.07 & 3560 & 0.34 & 0.04 & 3450 & 0.30 \\
J04180796+2826036 & M6 & 11.54 & 10.82 & 10.45 & 0.27 & 2990 & 0.073 & 0.37 & 2900 & 0.088 \\
J04182909+2826191 & M1 & 14.90 & 11.63 & 9.94 & 6.94 & 3705 & 2.2 & 6.81 & 3500 & 1.7 \\
J04183030+2743208 & M5.5 & 11.89 & 11.27 & 11.01 & 0.21 & 3057 & 0.051 & 0.04 & 3050 & 0.049 \\
J04184023+2824245 & M4 & 13.64 & 10.96 & 9.69 & 5.50 & 3270 & 1.5 & 5.12$^{\mathrm{a}}$ & 3300 & 1.1 \\
J04185115+2814332 & M7.5 & 13.93 & 13.24 & 12.75 & 0.34 & 2795 & 0.0088 & 0.25 & 2350 & 0.0072 \\
J04190197+2822332 & M5.5 & 11.99 & 10.78 & 10.15 & 1.72 & 3057 & 0.19 & 1.62 & 3050 & 0.19 \\
J04194127+2749484$^{\mathrm{d}}$ & M0 & 9.13 & 8.38 & 8.26 & 0.14 & 3850 & 0.87 & 0.02 & 3550 & 0.66 \\
J04203918+2717317 & M4.5 & 10.50 & 9.86 & 9.56 & 0.00 & 3197 & 0.16 & 0.11 & 3225 & 0.19 \\
J04205273+1746415 & M5.5 & 11.62 & 11.04 & 10.78 & 0.00 & 3057 & 0.054 & 0.00 & 3050 & 0.060 \\
J04214013+2814224 & M5.75 & 11.93 & 11.34 & 11.03 & 0.00 & 3023 & 0.040 & 0.05 & 3000 & 0.047 \\
J04215450+2652315 & M8.5 & 15.53 & 14.50 & 13.90 & 0.27 & 2555 & 0.0020 & 0.76 & 2100 & 0.0028 \\
J04220313+2825389$^{\mathrm{d}}$ & M3 & 9.46 & 8.67 & 8.45 & 0.14 & 3415 & 0.50 & 0.20 & 3375 & 0.55 \\
J04221332+1934392 & M8 & 12.86 & 12.05 & 11.52 & 0.00 & 2710 & 0.018 & 0.41 & 2225 & 0.022 \\
J04221644+2549118 & M7.75 & 13.06 & 12.36 & 11.94 & 0.14 & 2752 & 0.016 & 0.12 & 2275 & 0.014 \\
J04222404+2646258 & M4.75 & 11.09 & 10.19 & 9.77 & 0.27 & 3161 & 0.12 & 0.76 & 3200 & 0.20 \\
J04244506+2701447 & M5 & 11.34 & 10.71 & 10.46 & 0.14 & 3125 & 0.082 & 0.06 & 3150 & 0.083 \\
J04270739+2215037 & M6.75 & 12.27 & 11.65 & 11.29 & 0.07 & 2907 & 0.031 & 0.06 & 2800 & 0.032 \\
J04272799+2612052 & M9.5 & 15.00 & 14.02 & 13.28 & 0.00 & 2300 & 0.0026 & 0.43 & 1925 & 0.0040 \\
J04274538+2357243 & M8.25 & 14.93 & 14.24 & 13.69 & 0.00 & 2632 & 0.0026 & 0.17 & 2150 & 0.0027 \\
J04292071+2633406$^{\mathrm{d}}$ & M4 & 9.82 & 9.09 & 8.79 & 0.34 & 3270 & 0.43 & 0.24 & 3300 & 0.40 \\
J04292971+2616532$^{\mathrm{d}}$ & M5.5 & 10.34 & 9.68 & 9.39 & 0.27 & 3057 & 0.23 & 0.16 & 3050 & 0.22 \\
J04294247+2632493$^{\mathrm{d}}$ & M0 & 9.32 & 8.60 & 8.39 & 0.07 & 3850 & 0.68 & 0.13 & 3550 & 0.60 \\
J04294568+2630468 & M7.5 & 12.64 & 11.92 & 11.54 & 0.21 & 2795 & 0.026 & 0.14 & 2350 & 0.021 \\
J04300357+1813494 & M2 & 9.87 & 8.95 & 8.92 & 0.27 & 3560 & 0.45 & 0.08$^{\mathrm{a}}$ & 3450 & 0.34 \\
J04302365+2359129 & M8.25 & 14.96 & 14.24 & 13.70 & 0.00 & 2632 & 0.0026 & 0.20 & 2150 & 0.0027 \\
J04311578+1820072 & M4.25 & 11.21 & 10.55 & 10.30 & 0.07 & 3233 & 0.091 & 0.04 & 3250 & 0.093 \\
J04311907+2335047 & M7.75 & 13.51 & 12.72 & 12.19 & 0.14 & 2752 & 0.011 & 0.42 & 2275 & 0.012 \\
J04312382+2410529$^{\mathrm{d}}$ & M4.75 & 9.73 & 9.06 & 8.77 & 0.21 & 3161 & 0.39 & 0.16 & 3200 & 0.40 \\
J04312405+1800215 & M7 & 11.65 & 10.92 & 10.57 & 0.27 & 2880 & 0.066 & 0.20 & 2750 & 0.064 \\
J04312669+2703188 & M7.5 & 14.83 & 13.97 & 13.45 & 0.14 & 2795 & 0.0032 & 0.60 & 2350 & 0.0043 \\
J04315844+2543299 & M5.5 & 10.59 & 9.83 & 9.56 & 0.55 & 3057 & 0.23 & 0.28$^{\mathrm{a}}$ & 3050 & 0.20 \\
J04320329+2528078 & M6.25 & 11.72 & 11.11 & 10.72 & 0.00 & 2962 & 0.049 & 0.15 & 2875 & 0.060 \\
J04321786+2422149 & M5.75 & 11.54 & 10.79 & 10.38 & 0.00 & 3023 & 0.058 & 0.49 & 3000 & 0.10 \\
J04321885+2422271$^{\mathrm{d}}$ & M0.5 & 9.54 & 8.43 & 8.11 & 1.10 & 3777 & 1.41 & 0.94 & 3525 & 1.042 \\
J04322329+2403013 & M7.75 & 12.34 & 11.69 & 11.33 & 0.00 & 2752 & 0.028 & 0.00 & 2275 & 0.025 \\
J04322627+1827521 & M5.25 & 11.12 & 10.37 & 10.17 & 0.21 & 3091 & 0.11 & 0.15$^{\mathrm{a}}$ & 3125 & 0.11 \\
J04325026+2422115 & M7.5 & 13.96 & 12.22 & 11.28 & 3.16 & 2795 & 0.11 & 2.77$^{\mathrm{a}}$ & 2350 & 0.071 \\
J04325119+1730092 & M8.25 & 14.69 & 13.99 & 13.56 & 0.00 & 2632 & 0.0033 & 0.01 & 2150 & 0.0029 \\
J04330197+2421000 & M6 & 10.86 & 10.14 & 9.73 & 0.55 & 2990 & 0.18 & 0.43 & 2900 & 0.17 \\
J04330781+2616066 & M6 & 11.91 & 10.81 & 10.27 & 0.89 & 2990 & 0.093 & 1.27$^{\mathrm{a}}$ & 2900 & 0.14 \\
J04332621+2245293 & M4 & 11.80 & 10.50 & 9.92 & 1.10 & 3270 & 0.14 & 1.66$^{\mathrm{b}}$ & 3300 & 0.24 \\
J04334291+2526470 & M8.75 & 14.64 & 13.85 & 13.33 & 0.00 & 2477 & 0.0035 & 0.15 & 2050 & 0.0037 \\
J04335252+2256269 & M5.75 & 10.24 & 9.47 & 9.11 & 0.00 & 3023 & 0.19 & 0.45 & 3000 & 0.32 \\
J04335546+1838390$^{\mathrm{d}}$ & M3.5 & 10.53 & 9.87 & 9.61 & 0.14 & 3342 & 0.18 & 0.04 & 3325 & 0.18 \\
J04341099+2251445 & M1 & 10.59 & 9.74 & 9.43 & 0.48 & 3705 & 0.29 & 0.46 & 3500 & 0.25 \\
J04341527+2250309 & M7 & 13.74 & 12.54 & 11.85 & 1.65 & 2880 & 0.034 & 1.55 & 2750 & 0.033 \\
J04344544+2308027 & M5.25 & 12.81 & 12.02 & 11.70 & 0.48 & 3091 & 0.029 & 0.43$^{\mathrm{a}}$ & 3125 & 0.030 \\
J04345693+2258358 & M1.5 & 10.47 & 9.59 & 9.27 & 0.55 & 3632 & 0.34 & 0.51 & 3475 & 0.30 \\
J04350850+2311398 & M6 & 12.53 & 11.94 & 11.59 & 0.00 & 2990 & 0.023 & 0.10 & 2900 & 0.028 \\
J04352450+1751429$^{\mathrm{d}}$ & M2 & 10.03 & 9.33 & 9.08 & 0.14 & 3560 & 0.34 & 0.08 & 3450 & 0.30 \\
J04354183+2234115 & M5.75 & 12.95 & 12.37 & 11.98 & 0.14 & 3023 & 0.018 & 0.14 & 3000 & 0.020 \\
J04354203+2252226 & M4.75 & 11.25 & 10.39 & 9.99 & 0.48 & 3161 & 0.12 & 0.67 & 3200 & 0.16 \\
J04354526+2737130 & M9.25 & 15.02 & 14.24 & 13.71 & 0.00 & 2350 & 0.0026 & 0.00 & 1975 & 0.0024 \\
J04355109+2252401 & M2.75 & 11.31 & 10.35 & 10.01 & 0.55 & 3451 & 0.14 & 0.67 & 3400 & 0.16 \\
J04355143+2249119 & M8.5 & 15.48 & 14.66 & 14.19 & 0.48 & 2555 & 0.0025 & 0.22 & 2100 & 0.0018 \\
J04355209+2255039 & M4.5 & 11.31 & 10.23 & 9.81 & 0.55 & 3197 & 0.13 & 1.07$^{\mathrm{c}}$ & 3225 & 0.22 \\
J04355286+2250585 & M4.25 & 10.99 & 10.11 & 9.75 & 0.34 & 3233 & 0.14 & 0.61 & 3250 & 0.19 \\
J04355892+2238353 & M0 & 9.32 & 8.60 & 8.37 & 0.14 & 3850 & 0.73 & 0.13 & 3550 & 0.61 \\
J04361038+2259560 & M7.5 & 13.75 & 12.76 & 12.17 & 0.55 & 2795 & 0.013 & 0.93 & 2350 & 0.016 \\
J04361909+2542589 & M0 & 9.34 & 8.71 & 8.58 & 0.00 & 3850 & 0.63 & 0.00 & 3550 & 0.53 \\
J04363893+2258119 & M7.75 & 13.72 & 12.86 & 12.37 & 0.48 & 2752 & 0.012 & 0.51 & 2275 & 0.011 \\
J04380083+2558572 & M7.25 & 11.54 & 10.62 & 10.10 & 0.96 & 2837 & 0.14 & 0.75 & 2650 & 0.11 \\
J04383528+2610386$^{\mathrm{d}}$ & M1 & 9.23 & 8.28 & 7.91 & 0.76 & 3705 & 1.33 & 0.71 & 3500 & 1.1 \\
J04400174+2556292 & M5.5 & 13.22 & 11.64 & 10.76 & 1.44 & 3057 & 0.047 & 2.66$^{\mathrm{b}}$ & 3050 & 0.16 \\
J04410424+2557561 & M5 & 10.95 & 10.26 & 9.95 & 0.34 & 3125 & 0.14 & 0.25 & 3150 & 0.14 \\
J04414565+2301580$^{\mathrm{d}}$ & M4.5 & 10.74 & 10.10 & 9.85 & 0.14 & 3197 & 0.15 & 0.03 & 3225 & 0.14 \\
J04420548+2522562$^{\mathrm{d}}$ & M0 & 9.79 & 8.66 & 8.23 & 1.24 & 3850 & 1.31 & 1.16 & 3550 & 1.0 \\
J04420732+2523032$^{\mathrm{d}}$ & M1 & 9.58 & 8.40 & 7.95 & 1.37 & 3705 & 1.70 & 1.24 & 3500 & 1.3 \\
J04464260+2459034$^{\mathrm{d}}$ & M4 & 11.26 & 10.67 & 10.34 & 0.00 & 3270 & 0.082 & 0.05 & 3300 & 0.090 \\
J04484189+1703374 & M7 & 13.52 & 12.93 & 12.49 & 0.00 & 2880 & 0.0092 & 0.12 & 2750 & 0.011 \\
J04552333+3027366 & M6.25 & 13.06 & 12.38 & 11.97 & 0.41 & 2962 & 0.021 & 0.32 & 2875 & 0.020 \\
J04554046+3039057 & M5.25 & 12.72 & 12.07 & 11.77 & 0.07 & 3091 & 0.021 & 0.15 & 3125 & 0.025 \\
J04554757+3028077 & M4.75 & 11.05 & 10.31 & 9.98 & 0.00 & 3161 & 0.095 & 0.35 & 3200 & 0.14 \\
J04554820+3030160 & M4.5 & 11.89 & 11.22 & 10.95 & 0.00 & 3197 & 0.045 & 0.12 & 3225 & 0.053 \\
J04555288+3006523 & M5.25 & 11.65 & 11.03 & 10.73 & 0.21 & 3091 & 0.065 & 0.09 & 3125 & 0.064 \\
J04555636+3049374 & M5 & 12.00 & 11.40 & 11.09 & 0.07 & 3125 & 0.042 & 0.11 & 3150 & 0.047 \\
J04574903+3015195 & M9.25 & 15.77 & 15.12 & 14.48 & 0.00 & 2350 & 0.0013 & 0.00 & 1975 & 0.0012 \\
J05061674+2446102$^{\mathrm{d}}$ & M4 & 10.79 & 10.09 & 9.81 & 0.27 & 3270 & 0.16 & 0.16 & 3300 & 0.15 \\
J05064662+2104296 & M5.25 & 12.05 & 11.41 & 11.11 & 0.14 & 3091 & 0.042 & 0.13 & 3125 & 0.046 \\
        \hline
\end{longtable}
\begin{minipage}{.9\linewidth}
\renewcommand{\footnoterule}{}
\footnotetext[1]{ Flagged due to the best fit \AJ \ having an rms above 1.00.}
\footnotetext[2]{ Flagged due to the best fit \AJ \ differing to the \citet{Luhman10} derived value by above 0.5.}
\footnotetext[3]{ Flagged for a combination of $^{\mathrm{a}}$ and $^{\mathrm{b}}$ as above}
\footnotetext[4]{ Known binary \citep{Kraus11}.}
\end{minipage}
\normalsize

\begin{longtable}{lllll}
\caption[Luhman and Allard Temperature Scales and Bolometric Corrections]{Luhman and Allard Temperature Scales and Bolometric Corrections} \label{LTvsAT} \\
\hline \hline \multicolumn{1}{c}{Spectral Type} & \multicolumn{1}{c}{\LT(K)} & \multicolumn{1}{c}{\AT(K)} & \multicolumn{1}{c}{$BC_{J,L}$} & \multicolumn{1}{c}{$BC_{J,A}$} \rule{0pt}{2.6ex} \rule[-1.2ex]{0pt}{0pt} \\ \hline 
\endfirsthead
    M0            & 3850 & 3550 & 1.52 & 1.71 \\
    M1            & 3705 & 3500 & 1.58 & 1.72 \\
    M2            & 3560 & 3450 & 1.64 & 1.73 \\
    M3            & 3415 & 3375 & 1.79 & 1.75 \\
    M4            & 3270 & 3300 & 1.81 & 1.76 \\
    M5            & 3125 & 3150 & 1.88 & 1.79 \\
    M6            & 2990 & 2900 & 1.93 & 1.83 \\
    M7            & 2880 & 2750 & 1.93 & 1.89 \\
    M8            & 2710 & 2225 & 1.89 & 2.04 \\
    M9            & 2400 & 2000 & 1.84 & 1.93 \\
    L0            & 2200 & 1875 & 1.79 & 1.67 \\
    \hline
\end{longtable}

\end{document}